\documentclass[eqsecnum,amsmath,preprintnumbers,superscriptaddress,nofootinbib,aps,11pt]{revtex4} 

\usepackage{graphicx}
\usepackage{subfigure}
\usepackage{hyperref}
\usepackage{array}
\usepackage{amsmath}

\long\def\del #1 \enddel { }

\usepackage{graphicx}
\usepackage{amsmath}
\usepackage{amssymb}
\usepackage{subfigure}

\usepackage{epsfig}

\usepackage{graphicx}
\usepackage{subfigure}
\usepackage{hyperref}

\usepackage{color}

\def\beq{\begin{equation}}
\def\eeq{\end{equation}}

\def\bea{\arraycolsep .1em \begin{eqnarray}}
\def\eea{\end{eqnarray}}
\def\Tr{{\rm Tr}}

\def\eq#1{(\ref{#1})}

\def\R{\rho}
\def\R{R}
\def\s0#1#2{\mbox{\small{$ \frac{#1}{#2} $}}}
\def\0#1#2{\frac{#1}{#2}}

\def\grgl{\:\hbox to -0.2pt{\lower2.5pt\hbox{$\sim$}\hss}{\raise3pt\hbox{$>$}}\:}
\def\klgl{\:\hbox to -0.2pt{\lower2.5pt\hbox{$\sim$}\hss}{\raise3pt\hbox{$<$}}\:}

\begin{document}

\title{${}$\\[10ex] 
Further evidence for asymptotic safety of quantum gravity}
\author{K.~Falls}
\affiliation{Department of Physics and Astronomy, University of Sussex, Brighton, BN1 9QH, U.K.}
\affiliation{Institute for Theoretical Physics, U Heidelberg, 69120 Heidelberg, Germany}
\author{D.~Litim}
\author{K.~Nikolakopoulos}
\affiliation{Department of Physics and Astronomy, University of Sussex, Brighton, BN1 9QH, U.K.}
\author{C.~Rahmede}
\affiliation{{Karlsruhe Institute of Technology,  Institute for Theoretical Physics,  
76128 Karlsruhe, Germany}}
\affiliation{\mbox{Technical University Dortmund, Institute for Theoretical Physics,  
44221 Dortmund, Germany}}

\begin{abstract}
The asymptotic safety conjecture is examined for quantum gravity in four dimensions.
Using the renormalisation group, we find evidence for an interacting UV fixed point for polynomial actions up to the 34th power in the Ricci scalar. The extrapolation to infinite polynomial order is given, and the self-consistency of the fixed point is established using a bootstrap test. All details of our analysis are provided.  We also clarify further aspects such as stability, convergence, the role of boundary conditions, and a partial degeneracy of eigenvalues. Within this setting we find strong support for the conjecture.

\vskip14cm
{\noindent \footnotesize Preprint:  DO-TH 14/26, KA-TP-2014-30}

\end{abstract}

\maketitle
\newpage
\tableofcontents

\newpage
\section{Introduction}
This paper investigates the asymptotic safety conjecture for gravity, and is the continuation of a study initiated in \cite{Falls:2013bv}. 
Asymptotic safety for gravity stipulates that a fully-fledged quantum theory of the metric field may exist fundamentally, provided the short distance  fluctuations of the metric field lead to an interacting fixed point   \cite{Weinberg:1980gg}. The importance of  ultraviolet (UV) fixed points for the definition of quantum field theory has been noted long ago \cite{Wilson:1971bg,Wilson:1971dh}. Many  theories are known where the fixed point is non-interacting, a prominent example being asymptotic freedom of QCD  \cite{Gross:1973id,Politzer:1973fx}. Much less is known about the existence of interacting UV fixed points. A few rigorous results  are available in settings where perturbation theory remains intact. In 4D quantum gauge theories, interacting UV fixed points have been found recently in \cite{Litim:2014uca}. For gravity, reliable interacting UV fixed points arise close to two dimensions,  in the spirit of the epsilon expansion \cite{Gastmans:1977ad,Christensen:1978sc,Weinberg:1980gg,Kawai:1989yh,Kawai:1995ju}, or with the help of large-$N$ techniques \cite{Tomboulis:1977jk,Tomboulis:1980bs,Smolin:1981rm}, where $N$ denotes the number of matter fields. 

Identifying interacting UV fixed points in a reliable manner faces two key challenges. Firstly, if a theory is asymptotically free, the set of relevant, marginal, and irrelevant invariants is known beforehand. It then suffices 
to retain the  finite set of classically relevant and marginal invariants in the  action. Provided a theory is asymptotically safe, residual interactions in the UV modify the power counting of invariants. The set of relevant and marginal invariants is then no longer known beforehand, and it cannot be taken for granted that invariants which are classically irrelevant will remain irrelevant at an interacting fixed point \cite{Weinberg:1980gg}. For gravity, one may then wonder whether high powers of e.g.~the Ricci scalar, such as $R^{256}$, may become relevant at an interacting fixed point? The task therefore must consist in identifying a procedure by which a fixed point can be identified, self-consistently, despite of the fact that explicit studies are often confined to a finite number of invariants \cite{Falls:2013bv}. Secondly, in four-dimensional gravity, Newton's coupling carries inverse mass dimensions and conventional pertubation theory is not 
applicable at highest energies. Furthermore, the theory is not offering a natural small expansion parameter, and non-perturbative techniques 
are required to deal with strong coupling effects.

Interestingly, 
the lack of a priori information about the set of relevant invariants  can be compensated with the help of  an auxiliary hypothesis  \cite{Falls:2013bv}. We will assume that invariants with increasing canonical mass dimension  remain increasingly irrelevant at an interacting UV fixed point. The  rationale for this  relates to the fact  that quantum fluctuations would have to overcompensate  increasingly large canonical mass dimensions to turn irrelevant invariants into relevant ones \cite{Weinberg:1980gg}. It is then  conceivable that an ordering according to the canonical mass dimension remains a good principle even in the interacting quantum theory. The virtue of the auxiliary hypothesis is that it can be falsified, allowing for systematic tests of the asymptotic safety conjecture. 

In this paper, we  test the asymptotic safety conjecture for quantum gravity  
in concrete terms.  The primary questions we wish to address with this are:  Can an  interacting UV  fixed point be identified self-consistently, and if so, what are its properties? What is the impact of high-order curvature invariants?  Is it safe to assume that the canonical mass dimension offers a good guiding principle? We study these topics, examplarily,
for gravitational actions which are  high-order polynomials in the Ricci scalar. With Newton's coupling, the cosmological constant and the $R^2$ coupling, these models contain three of the classically relevant and marginal invariants, plus an increasing number of canonically irrelevant invariants.  Curvature invariants other than powers of the Ricci scalar are neglected, which is our main approximation.  We expect that an interacting UV fixed point, should it exist in the full theory, becomes visible even if only a subset of invariants is taken into account. As such, this paper is an extension of \cite{Falls:2013bv} including more background, details, and further insights.

For our explicit computations we adopt a functional version of Wilson's renormalisation group (RG) \cite{Wetterich:1992yh,Morris:1993qb}, which is based on the successive integrating-out of momentum degrees of freedom. A feature of this continuum method is that it can be applied even at strong coupling \cite{Berges:2000ew}. Optimisation techniques are available to maximise the physics content within given approximations, also offering analytical access to the relevant RG flows \cite{Litim:2000ci,Litim:2001up}. Furthermore, a large body of work exists showing that these techniques can  be used to access interacting fixed points and strong coupling effects \cite{Berges:2000ew,Litim:2010tt}. For gravity, these methods have been made available in \cite{Reuter:1996cp}, see \cite{Litim:2006dx,
Niedermaier:2006ns,
Litim:2008tt,
Litim:2011cp,Percacci:2011fr,Reuter:2012id} for reviews. Applications thus far  include Einstein-Hilbert, higher-derivative, non-local, and $f(R)$-type approximations 
\cite{Reuter:1996cp,Dou:1997fg,Souma:1999at,
Lauscher:2001ya,Lauscher:2002sq,Litim:2003vp,Fischer:2006at,Fischer:2006fz,Codello:2006in,Codello:2007bd,Codello:2008vh,
Machado:2007ea,Bonanno:2010bt,Benedetti:2012dx,Dietz:2012ic,Benedetti:2013jk,Christiansen:2014raa}.
To connect with some of the earlier work, we adopt the rationale of \cite{Litim:2003vp,Codello:2007bd,Codello:2008vh,Machado:2007ea}. The strength of this setup is that it admits a well-controlled heat kernel expansion in powers of the Ricci scalar. Our main technical novelty here is to provide ways how the expansion and the fixed point search can be performed to very high polynomial order by combining algebraic and numerical methods.

The outline of the paper is as follows.
 In Sec.~\ref{RG}., we recall the renormalisation group, introduce some notation,  and specify our approximations.
We then analyse the classical and quantum fixed points of our model (Sec.~\ref{FP}), provide a systematic algorithm to determine fixed point coordinates algebraically, and determine two remaining free parameters numerically, including error estimates. A stable fixed point is identified for polynomial actions in the Ricci scalar up to some maximum order $N=35$. The good convergence permits a large-$N$ extrapolation, the results of which are also given.
In Sec.~\ref{SE} we present our results for the scaling exponents, including a discussion of degenerate eigenvalues, the enhanced gap in the eigenvalue spectrum,  convergence, and an underlying eight-fold periodicity pattern. We also analyse the impact of the boundary condition on the fixed point search, finding that the convergence is improved through suitable choices (Sec.~\ref{Stability}). In Sec.~\ref{PC}, we review the bootstrap hypothesis for asymptotic safety, and explain how it is realised in the data. 
A brief discussion of the near-Gaussianity of subleading eigenvalues is given in   Sec.~\ref{NG}, followed by our conclusions (Sec.~\ref{C}). 
An appendix provides more details of the explicit RG equations (App.~\ref{AppA}).

\section{Gravitational renormalisation group}\label{RG}
In this section, we introduce our set-up and detail the relevant RG equations. We adopt the framework of the functional renormalisation group for gravity, which is based on a Wilsonian momentum cutoff to successively integrate-out momentum modes.
We begin with euclidean gravitational actions which are functions of the Ricci scalar $R$,
\beq\label{S}
S=\int d^4x \sqrt{\det g_{\mu\nu}}\,
F(R)\,.
\eeq
where $R=R(g_{\mu\nu})$ denotes the Ricci scalar and $g_{\mu\nu}$ the metric field.
Actions which are generic functions of the curvature scalar are of interest for cosmological model building and dark energy, see \cite{DeFelice:2010aj}. Classically, they can be re-written as standard Einstein-Hilbert gravity coupled to a scalar field with a potential determined through the function $F$. Expanding the action polynomially, we recover the Einstein-Hilbert action
\beq
F(R)=\frac{\Lambda}{8\pi\,G}-\frac{1}{16\pi\,G} R+\cdots\,.
\eeq
up to higher order corrections in the Ricci scalar. Here, $\Lambda$ denotes the cosmological constant, $G=6.67 \times  10^{−11}\,{\rm  m}^3/({\rm kg\, s}^2)$
 Newton's constant, and ${\Lambda}/(8\pi\,G)$ the vacuum energy. In general, the action \eq{S} need not to be polynomial in the curvature scalar. 

We are specifically interested in the quantum theory associated to a polynomial action in the Ricci scalar
 in view of the asymptotic safety conjecture for gravity. Provided the theory develops a non-trivial UV fixed point, it may become a candidate for a fundamental quantum theory of the metric field \cite{Weinberg:1980gg}. As soon as quantum fluctuations are taken into account, the couplings turn into running couplings  $\Lambda\to \Lambda_k$, $G\to G_k$ and  $F\to F_k$, whereby the classical action \eq{S} becomes a quantum effective action $\Gamma_k$ to evolve with the RG momentum scale $k$ at which the theory is probed. 
 A particularly useful continuum method to describe the change of the gravitational effective action with RG momentum scale is given by Wilson's renormalisation group. It is based on a coarse-grained version of the path integral where the propagation of fluctuations with momenta smaller than the RG scale $k$ are suppressed. In its modern form, the dependence of the effective action on the RG scale is  given by an exact functional identity \cite{Wetterich:1992yh}
\beq\label{FRG}
\partial_t\Gamma_k=\frac12{\rm S}\Tr\frac{1}{\Gamma^{(2)}_k+R_k}\partial_t R_k\,,
\eeq
and $t=\ln (k/k_0)$ where $k_0$ is some arbitrary reference scale which does not enter into any of the later results. Here, the (super)trace stands for a sum over modes and fields including appropriate minus signs for fermions and ghosts. The regulator function $R_k$ can be chosen at will, though within a few constraints to ensure that the RG flow interpolates between the microscopic theory in the UV and the full quantum effective theory in the IR. We exploit this freedom to obtain explicit analytical RG flows for all couplings  \cite{Litim:2003vp} adopting the ideas of \cite{Litim:2000ci,Litim:2001up}. 

Functional flows \eq{FRG} for actions \eq{S} have been derived in \cite{Codello:2007bd,
Codello:2008vh,Machado:2007ea}, and in \cite{Benedetti:2012dx} based on the on-shell action, also using \cite{Litim:2001up,Litim:2003vp}.
Diffeomorphism symmetry is controlled with the help of the background field method \cite{Reuter:1996cp,Freire:2000bq} which splits the metric $g_{\mu\nu}=\bar{g}_{\mu\nu}+h_{\mu\nu}$ into a classical background $\bar{g}_{\mu\nu}$ and a quantum part $h_{\mu\nu}$. For $F(R)$ theories it is sufficient to choose the background metric to be that of a maximally symmetric four-sphere to obtain a closed flow equation. This is achieved by expanding $\Gamma_k$ to quadratic order in the fluctuation $h_{\mu \nu}$ around the four sphere, taking the second variation to obtain $\Gamma^{(2)}_k$ and then  setting $g_{\mu \nu} = \bar{g}_{\mu \nu}$. To facilitate consistency checks and a comparison with earlier findings we have adopted the approach put forward in \cite{Codello:2007bd,Codello:2008vh} with the same choice of gauge-fixing
\begin{equation}
\label{fofrgaugefixing}
S_{GF}=\frac{1}{2\alpha}\int d^4 x\sqrt{\bar
g}\,\chi_{\mu}
\chi^{\mu}
\end{equation}
where
$\chi_{\nu}=\bar{\nabla}^{\mu}h_{\mu\nu}-\frac{1}{4}\bar{\nabla}_{\nu}h^{\mu}_{\,\,\mu}$
which leads to a ghost part of the action for the ghost field $C_{\mu}$
\begin{equation}
S_{gh}=\frac{1}{\alpha}\int d^4 x\sqrt{\bar g}\,{\bar C}_{\nu}
\frac{\delta \chi^{\nu}}{\delta
\epsilon^{\mu}}C^{\mu}.
\end{equation}
In this computation we will use the Landau-De Witt gauge $\alpha\to 0$, 
which simplifies the flow equation.
It has been shown \cite{Litim:1998qi} that this value of $\alpha$ is a fixed point value for the running gauge parameter.

In order to evaluate the trace via heat kernel techniques we need to have the second variation organised in terms of the laplacian operator $\nabla^2$ on the four sphere \footnote{From here on we drop the bar notation, where it is understood that all metrics, covariant derivatives etc. are evaluated on the maximally symmetric background, e.g. $\nabla^2 = \bar{\nabla}^2$} . For this reason we decompose the quantum fluctuations, according to the transverse-traceless decomposition \cite{York:1973ia} which was first used for RG computations in \cite{Dou:1997fg}
\begin{equation}
\label{decomposition}
h_{\mu\nu}=h^T_{\mu\nu}+\nabla_{\mu}\xi_{\nu}
+\nabla_{\nu}\xi_{\mu}+\nabla_{\mu}\nabla_{\nu}\sigma-\frac{1}{
4}g_{\mu\nu}\nabla^2\sigma+\frac{1}{4}g_{\mu\nu}h.
\end{equation}
Here $h=g^{\mu\nu}h_{\mu\nu}$ is the trace of the fluctuation, $h_{\mu\nu}^T$ denotes the transverse-traceless part of $h_{\mu\nu}$, $\xi_{\mu}$ is a transverse vector that together with the scalar $\sigma$ makes up the longitudinal-traceless part of $h_{\mu\nu}$ according to \eqref{decomposition}. These fields obey the differential constraints:
\begin{equation}
h_{\mu}^{T\mu}=0\ ; \qquad\nabla^{\nu}h_{\mu\nu}^{T}=0\ ;
\qquad\nabla^{\nu}\xi_{\nu}=0\ .
\end{equation}
The advantage of this decomposition, along with the gauge choice $\alpha\to 0$, is that the RG flow \eq{FRG} simplifies considerably, splitting into a sum of several traces. Those over $h_{\mu\nu}^T$ and $h$ are independent of the gauge parameter and come only from $F_k(R)$. In the traces containing contributions from $\xi$ and $\sigma$, the gauge fixing term dominates since these traces become independent of $F_k(R)$. 

Similarly, for the ghost fields we adopt their decomposition into transverse and longitudinal parts according to
\begin{equation}
\label{ghostdecomposition}
C^{\mu}=c^T{}^{\mu}+\nabla^{\mu} c\ \ , {\bar C}_{\mu}={\bar
c}^T_{\mu}+\nabla_{\mu} {\bar c}\ ,
\end{equation}
where $c_{\mu}^T$ and $\bar c_{\mu}^T$ are transverse vectors and $c$ and $\bar c$ are scalars. They obey the differential constraints:
\begin{equation}
\qquad \nabla^\mu \bar c^{T}_{\mu}=0\  ;
\qquad\nabla_\mu c^{T\mu}=0\ .
\end{equation}
Since this decomposition involves a change of variables, it induces Jacobians for the transformation which appear as  determinants of the operators
\begin{equation}
J_V = -\nabla^2-\frac{R}{4},\,\, J_S =
-\nabla^2\left(-\nabla^2-\frac{R}{3}\right),\,\, J_c = -\nabla^2
\end{equation}
originating from the vector, scalar and ghost fields, respectively.
These can be properly taken into account by exponentiating the determinants with the introduction of some auxiliary field variables. The contributions resulting from the $\xi$ and $\sigma$ components together with the contributions from ghost fields and Jacobians simplify significantly.

The Wilsonian momentum cutoff function $R_k$ in \eqref{FRG} is chosen according to the prescription that for each individual component of the inverse propagator  $\Gamma^{(2)}_k$ we perform the substitution $-\nabla^2 \to -\nabla^2+\mathcal{R}_k(-\nabla^2)$ to obtain $\Gamma^{(2)}_k + R_k$, where $\mathcal{R}_k$ is the scalar cutoff function. For our computation we use the optimised cutoff \cite{Litim:2000ci,Litim:2001up,Litim:2003vp} given by
\begin{equation}\label{opt}
\mathcal{R}_k(y)=(k^2-y)\theta(k^2-y).
\end{equation}
RG flows with \eq{opt} are known to have good stability and convergence properties  \cite{Litim:2001fd,Litim:2001dt,Litim:2002cf,
Litim:2003vp,Fischer:2006at,Litim:2007jb,Litim:2010tt}. Equally important, the choice \eq{opt} 
also allows for analytical RG equations \cite{Litim:2003vp}, which is central for our work.

For fixed point studies, it is convenient to introduce dimensionless couplings. The dimensionless Newton coupling $g$ 
and dimensionless cosmological constant $\lambda$ are defined as
\beq\label{gla}
g\equiv G_k\,k^2\,,\quad\lambda\equiv\frac{\Lambda_k}{k^2}\,.
\eeq
We also find it convenient to introduce a dimensionless Ricci scalar. From now on, for notational simplicity and unless stated otherwise, we denote it again as $R$ meaning that $
\bar R=R\,k^2$ instead refers to the dimensionful Ricci scalar. We then also introduce the dimensionless function $f(\R)$ as
\beq\label{f}
\frac{F_k(\bar R)}{k^4}=\frac{1}{16\pi}f(\R)
\eeq
where it is understood that $f$ is still a function of $k$. The factor $1/(16\pi)$  is purely conventional and has been adopted to ensure that the dimensionless Newton coupling is related to $f$ as $g=-1/f'(\R=0)$ without further numerical factors, see \eq{gla}. In general, the functional RG 
flow for \eq{f} takes the form
\beq\label{df}
\partial_t f + 4 f - 2\R\,f'=I[f]\,.
\eeq
The terms on the LHS account for the canonical running of couplings, and those on the RHS  originate from quantum fluctuations. 
In our case, the function $I[f]$ (given in appendix~\ref{AppA}) has homogeneity degree zero in $f$ with $I[a\,f]=I[f]$ for any $a\neq 0$. 
 Furthermore, the terms on 
the RHS also involve the flow of higher order derivatives of $f$ up to the second order,
\beq\label{I}
I[f]=I_0[f]+I_1[f]\cdot \partial_t f'+ I_2[f]\cdot \partial_t f''\,.
\eeq
This structure comes about due to background field dependences introduced via the Wilsonian regularisation \cite{Litim:2002cf,Litim:2002hj}, 
and also appears in (generalized) proper-time RG flows \cite{Litim:2002xm}. Additional flow terms on the RHS are expected to enhance the stability of the RG flow, as they correspond to effective resummations  \cite{Litim:2002hj}. The functions $I_n$ depend explicitly on $f$ and its first three derivatives, and on $\R$. Explicit expressions are given in the appendix~\ref{AppA}. Below, we exploit the RG flow \eq{df} as a generating function for the RG flows for all polynomial couplings of the theory.

\section{Fixed points}\label{FP}
In this section, we discuss classical and quantum fixed points, detail our numerical methods, and summarize results for a non-trivial ultraviolet fixed point. 

\subsection{Classical fixed points}

As a warm-up we first discuss the `classical' fixed points of our theory, as these may be achieved as asymptotic limits of the quantum theory. In the absence of fluctuations the RG flow \eq{df} becomes
\beq\label{classical}
\left(\partial_t  + 4  - 2\R\,\partial_\R\right)f=0\,.
\eeq
It states that all (dimensionful) couplings in the classical theory are independent of the energy scale. The linearity of the flow in $f$ implies the existence of a Gaussian fixed point 
$f_*\equiv 0$\,.
From the flow for the inverse 
\beq\label{classicalinv}
\left(\partial_t - 4- 2\R\,\partial_\R\right)( f^{-1})=0
\eeq
we also conclude the existence of an `infinite' Gaussian fixed point  \cite{PhysRevLett.32.1446} associated to the asymptotic vanishing of 
\beq
\label{infG}
1/f_*\equiv 0\,.
\eeq
More specifically, the RG flow \eq{classical} 
has the general solution
\beq\label{FPclassical}
f(\R,t)=\R^2 \cdot H\left(\R e^{2t}\right)
\eeq
for arbitrary function $H(x)$ which is determined by the boundary conditions at $t=0$. Fixed points correspond to $t$-independent solutions to \eq{FPclassical}. A trivially $t$-independent solution is achieved via the boundary condition $H(x)=$~const. It leads to a line of  fixed points corresponding to $R^2$-theories of gravity,
\beq\label{quad}
f_*=\lambda_2\,\R^2\,,
\eeq 
parametrized by the free parameter  $\lambda_2$, which has the role of a marginal coupling due to the vanishing canonical mass dimension of the $R^2$ coupling in four space-time dimensions. As such \eq{quad} is both an UV and IR fixed point. The Gaussian and infinite Gaussian fixed points arise from \eq{FPclassical} in asymptotic UV and IR  limits where $t\to\pm\infty$. The discussion of these cases is simplified due to the linearity of \eq{classical} and \eq{classicalinv}, and we can limit ourselves to the scaling analysis for monomials in the Ricci scalar $f\sim \lambda_n\R^n$ (no sum). The result \eq{FPclassical} then  states that the 
couplings 
scale canonically with Gaussian eigenvalues $\vartheta_{{\rm G}}$,
\beq\label{classicalscaling}
\begin{array}{rl}
\lambda_n(t)&=\lambda_n(0)\exp(\vartheta_{{\rm G},n} t)\\[1ex]
\vartheta_{{\rm G},n}&=2n-4\,.
\end{array}
\eeq
Consequently, the dimensionless vacuum energy term $(n=0)$ and the dimensionless Ricci coupling $(n=1)$ are relevant operators, and their dimensionless couplings diverge towards the IR, leading to the infinite Gaussian fixed point \eq{infG}. Using \eq{f}, we can relate the IR diverging couplings $\lambda_0$ and $\lambda_1$ to the dimensionless Newton coupling and cosmological constant to find $g\equiv -1/\lambda_1$ and   $\lambda\equiv - (\lambda_0)/(2\lambda_1)$, which 
translates into
\beq\label{GaussCosmo}
1/\lambda\to 0\,,\quad g\to 0\,.
\eeq
We conclude that general relativity with positive (negative) vacuum energy  corresponds to the  IR fixed point \eq{GaussCosmo}, provided that $\lambda$ is positive (negative). Furthermore, this fixed point is IR attractive in both couplings. The theory also displays an IR fixed point corresponding to a vanishing vacuum energy,
\beq\label{Gauss0}
\lambda=0\,,\quad g\to 0\,.
\eeq
This fixed point is IR attractive in $g$ and IR repulsive in $\lambda$, in contrast to \eq{GaussCosmo}. Classically, it can only be achieved by fine-tuning the vacuum energy to zero through the boundary condition.
This analysis can straightforwardly be extended to higher order monomials  including non-local ones, such as inverse powers in the Ricci scalar. 
According to \eq{classicalscaling}, for all couplings with $n>2$ $(n<2)$ the Gaussian fixed point $\lambda_n \to 0$ is IR attractive (repulsive) and therefore approached in the IR limit (UV limit), whereas the infinite Gaussian fixed point $1/\lambda_n \to 0$ is IR repulsive (attractive) and therefore approached in the UV limit (IR limit).

Next we discuss in which limits the classical fixed points may arise out of the full RG flow \eq{df}. To that end, we divide \eq{df} by $f$, finding
\beq\label{dfdimless}
4+\Big(\partial_t   - 2\R\,\partial_\R\Big)\,\ln f=I[f]/f\,.
\eeq
Note that the LHS of \eq{dfdimless} and $I[f]$ both have homogeneity degree zero in $f$. Furthermore, the fluctuation-induced term $I[f]$ is generically non-zero also in the limit of vanishing $f$.  However, the classical limit requires the vanishing of the RHS which, therefore, is parametrically achieved as
\beq\label{classlimit}
I[f]/f\to 0\quad{\rm for}\quad 1/f\to 0\,.
\eeq 
We thus conclude that the classical limit \eq{classlimit} arises from the full RG flow \eq{df} through the infinite Gaussian fixed point \eq{infG}. This specifically includes the IR fixed point for the couplings $\lambda_0$ and/or $\lambda_1$ which entail classical general relativity in the deep IR with a vanishing or non-vanishing  vacuum energy, see \eq{GaussCosmo}, \eq{Gauss0}. It also includes the possibility for a classical limit arising through \eq{quad} for asymptotically large-fields $1/\R\to0$, leading to an $R^2$-type theory. These results are straightforwardly extended to dimensions different from $d=4$.

\subsection{Strategy for quantum fixed points}
Next we turn to the fluctuation-induced fixed points of the theory, which arise through the non-vanishing RHS of \eq{df}. Provided that the RG flow \eq{f} has a non-trivial fixed point where $\partial_t f_*\equiv 0$, its location is determined by the function $I_0$,
\beq\label{fp}
0=- 4 f_* + 2\R\,f_*'+I_0[f_*]\,,
\eeq
see \eq{I}, and \eq{I0} for an explicit expression. A non-trivial UV fixed point is a candidate for an asymptotically safe short distance theory of gravity. An analytical solution for the third-order non-linear differential equation \eq{fp} is presently not at hand, and we have to content ourselves with approximate ones. To that end we adopt two complementary methods which have been tested successfully in critical scalar theories.

Firstly, we assume that the fixed point solution is polynomially expandable to high order, at least for small curvature scalar. If so, the fixed point condition provides equations for the polynomial couplings, which can be solved algebraically for all but a few couplings  \cite{Litim:2002cf}. Its solution constitutes a formally exact solution to \eq{fp} up to the highest order of the polynomial approximation, and within the radius of convergence of the expansion. The remaining free parameters must then be determined by other means, for example by imposing boundary conditions for the highest couplings. This corresponds to a bootstrap. The strength of this procedure is its algebraic exactness, leading to a maniable set of equations which can be extended systematically to higher orders. Furthermore, the expansion is best in the regime where the reliability of the heat kernel techniques used in the derivation of the flow are best. Finally, fixed points and universal exponents can reliably be deduced within a polynomial approximation \cite{Aoki:1998um,Litim:2002cf}. On the other hand, the weakness of our method is that a closure of the procedure requires certain assumptions about the highest couplings. The stability of a solution together with the boundary condition then needs to be tested with increasing polynomial order. Also, polynomial approximations are limited to a finite region in field space due to a finite radius of convergence. 

Extending polynomial fixed point solutions beyond this limit requires extra work. Here, we use direct numerical integration techniques to find the fixed point solution of \eq{fp}, without primarily relying on a polynomial approximation~\cite{Morris:1994ki,Litim:2002cf,Bervillier:2007rc}.
The strength of this strategy is that it makes no assumptions as to the functional form of its solution, polynomial or otherwise. In turn, the weakness of this procedure is that a numerical integration requires high-accuracy initial data, eg.~the derivatives of $f$ at vanishing curvature scalar. Also, the accuracy in the result is limited by that of the integration algorithm. Furthermore, identifying the fixed point for all fields may be hampered by technical artefacts for intermediate or large curvature scalar~\cite{Dietz:2012ic}.  Below, we combine both of these methods to test the reliability in our results.

\subsection{Algebraic fixed points}
We now discuss the algebraic procedure leading to closed expressions for the fixed point coordinates \cite{Litim:2002cf}. Our strategy is independent of the actual RG flow and can be adopted for other forms of the equation as well. We begin with a polynomial expansion of \eq{fp} about vanishing curvature scalar,
\beq\label{expansion0}
f(\R)=
\sum_{n=0}^{\infty}
\lambda_n\R^n\,.
\eeq
Inserting \eq{expansion0} into \eq{fp} leads to algebraic equations amongst all couplings. Specifically, the $\beta$-functions for all couplings follow from inserting \eq{expansion0} into \eq{df},
\begin{equation}\label{betan}
\beta_n\equiv\partial_t \lambda_n\,.
\end{equation}
The fixed point conditions $\beta_n=0$ can then be solved algebraically, order by order, starting at $n=0$. Evidently, these solutions are also solutions to \eq{fp}. Note that the differential equation \eq{df} with \eq{I} serves as a generating function for the $\beta$-functions of all polynomial couplings of the theory.  Solving $\beta_n=0$ starting with $n=0$ only constitutes definite equations for all but a finite set of couplings. The reason for this  is that the RG flow for a coupling $\lambda_n$ depends on the couplings up to $\lambda_{n+2}$. Therefore solving $\beta_n$ provides us with an expression for $\lambda_{n+2}$, and $\lambda_0$ and $\lambda_1$ remain unspecified to any order.
Hence, solving $\beta_{n-2}=0$ allows us to express $\lambda_n$ for all $n\ge 2$  in terms of lower order couplings, 
\begin{equation}
\lambda_n=\lambda_n(\lambda_i,i< n)\,.
\end{equation}
In a second step, these expressions for the couplings are further reduced, recursively, to functions  of the two unspecified couplings $\lambda_0$ and $\lambda_1$ only. This procedure provides us with an exact two-parameter family of fixed point candidates
\begin{equation}\label{algebraic}
\lambda_n=\lambda_n(\lambda_0,\lambda_1)\,,
\end{equation} 
where $n\geq 2$. For example, the coupling $\lambda_2$ is given by
\beq\label{lambda2}
\lambda_2(\lambda_0,\lambda_1)=-\frac{1}{9}\frac{12 \pi  \lambda _0^3+6 \left(5 \pi  \lambda _1+1\right) \lambda _0^2+2 \lambda _1 \left(9 \pi  \lambda _1+1\right)
   \lambda _0-9 \lambda _1^2}{12 \pi  \lambda _0^2+3 \left(4 \pi  \lambda _1+1\right) \lambda _0-7 \lambda _1}
\eeq
Similar, though increasingly more complex expressions are found for the higher order couplings.

Given the algebraic expressions \eq{algebraic}, it remains to identify the correct values for the remaining couplings $\lambda_0$ and $\lambda_1$, which are not determined by the algebraic procedure. To that end, we adopt the following strategy: we assume that a finite order approximation of \eq{expansion0} retaining the first $N$ couplings is a valid approximation. This implies that the couplings $\lambda_N$ and $\lambda_{N+1}$ no longer appear on the level of the action. We therefore may impose an auxiliary condition for the (unspecifed) higher-order couplings $\lambda_N$ and $\lambda_{N+1}$. Most of the times, we are adopting free boundary conditions,
\beq\label{boundary}
\begin{split}
\lambda_N        \, &=\ 0\\
\lambda_{N+1} \,&=\ 0\ .
\end{split}
\eeq
This boundary condition assumes that the couplings and the corresponding invariants are absent throughout, and that the recursive fixed point solution \eq{algebraic} should reflect this. This strategy has been tested previously for critical scalar theories. In practice, \eq{boundary} must be seen as an additional input into the search strategy, and its applicability needs to be confirmed a posteriori. We defer a detailed discussion of more general boundary conditions, and the stability of fixed point solutions, to Sec.~\ref{Stability}. 

At order $N$ in the approximation, each of the conditions \eq{boundary} with \eq{algebraic} leads to a constraint in the $(\lambda_0,\lambda_1)$ plane. Since the higher-order couplings are algebraic 
functions of $(\lambda_0,\lambda_1)$, the boundary conditions \eq{boundary} lead to a high-order polynomial equation in $\lambda_0$ (or $\lambda_1$).  In principle, these may have many roots in the complex plane. It then remains to identify those roots 
\beq\label{root}
\begin{split}
\lambda_0        \, &= \lambda_{0,*}\\
\lambda_{1} \,&= \lambda_{1,*}
\end{split}
\eeq
which are real, and numerically stable under extended approximations with increasing order $N$. If so, the fixed point qualifies as a candidate for a fundamental fixed point of the theory.

Polynomial expansions are not bound to the form \eq{expansion0} and can equally be performed about non-vanishing dimensionless Ricci scalar,
\beq
f(\R)=\sum_{n=1}^\infty\lambda_n(\R-\R_0)^n\,,
\eeq 
where $\R_0\neq 0$ is the expansion point. One finds that  all higher order couplings $\lambda_n\equiv n!\,f^{(n)}(\R_0)$ for $n>2$ can be expressed as rational functions in terms of three independent couplings $\lambda_0, \lambda_1$ and $\lambda_2$, except for a few exceptional points in field space where the recursive solution reduces to two independent couplings. Generically, three additional conditions are required to uniquely identify the fixed point. We have confirmed that this method works, but it is often more demanding than \eq{expansion0} to which we stick for most of our analysis.

\begin{figure}[t]
\centering
\begin{center}
\unitlength0.001\hsize
\includegraphics[width=.7\hsize]{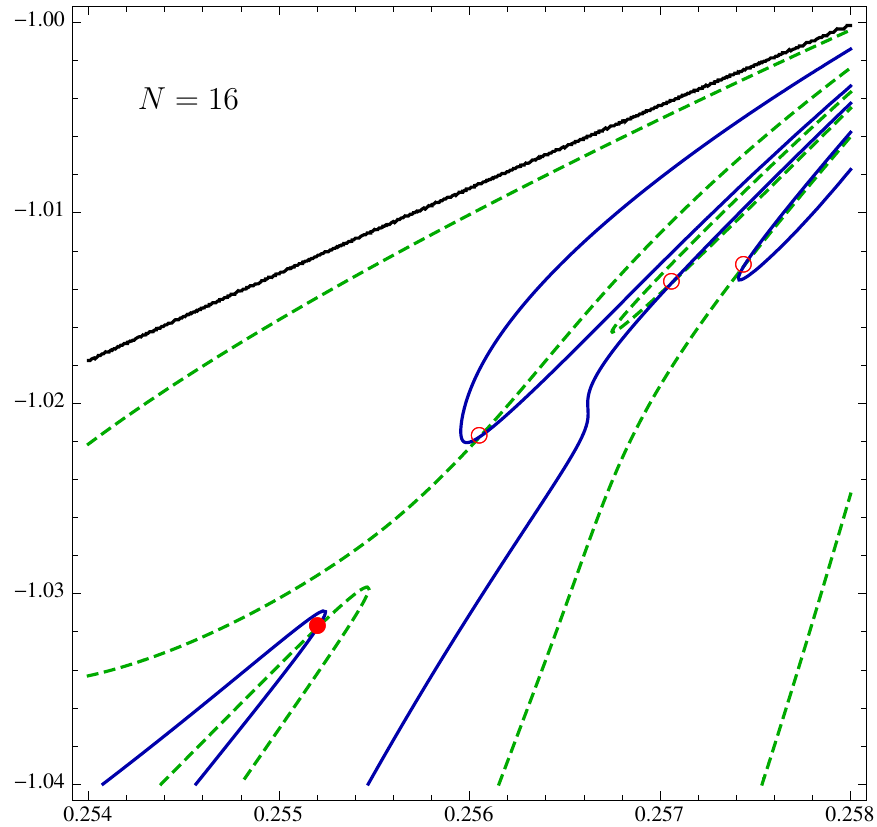}
\caption{\label{Contour16} Contour plot for the fixed point condition at order $N=16$ in the $(\lambda_0,\lambda_1)$ plane. Shown are the nullclines $P_{16}=0$ (full blue line) and $P_{17}=0$ (dashed green line) as well as the nullcline $Q_{16}=0$ (full black line). The nullcline $Q_{17}=0$ is outside the plotted region. Consistency conditions \eq{C1}, \eq{C2} identify the lower left fixed point, indicated by a full red circle, as a reliable candidate. Three empty red circles indicate  fixed point candidates which have failed the consistency condition \eq{C2} (see text).} 
\end{center}
\end{figure}

\begin{figure}[t]
\centering
\begin{center}
\unitlength0.001\hsize
\includegraphics[width=.48\hsize]{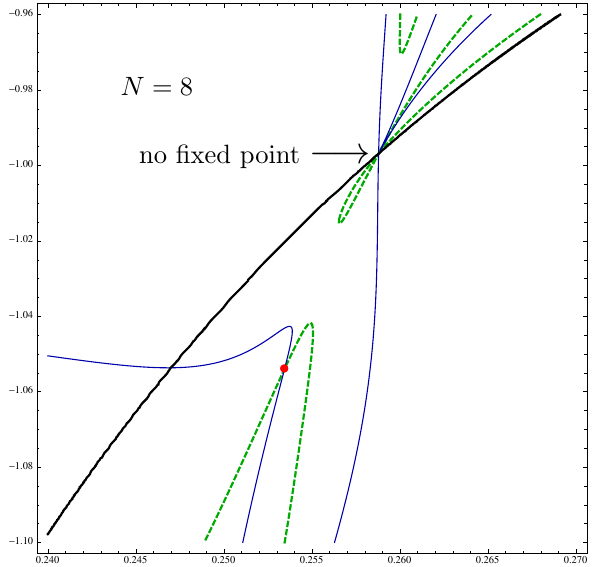}
\includegraphics[width=.48\hsize]{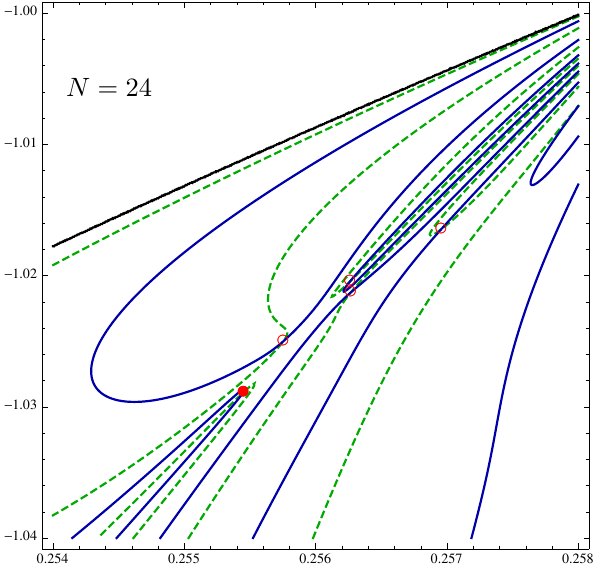}
\caption{\label{Contour8+24} Same as Fig.~\ref{Contour16}, but for the fixed point condition at approximation order $N=8$ (left panel) and $N=24$ (right panel). Shown are the nullclines $P_{8}=0$ and $P_{24}=0$ (full blue line),  $P_{9}=0$ and $P_{25}=0$ (dashed green line) as well as the nullclines $Q_{8}=0$ and $Q_{24}=0$ (full black line), respectively. The nullclines $Q_{9}=0$ and $Q_{25}=0$ are outside the plotted regions. In either case, consistency conditions identify the lower left fixed point (full red circle) as a reliable candidate. Four empty red circles in the right panel indicate  fixed point candidates which have failed the consistency test \eq{C2}. The left panel also shows an example where the nullclines $P_{8}=0=P_9$ have a joined zero with the nullcline $Q_9=0$ (no fixed point). Comparing with Fig.~\ref{Contour16} we note that the density of fixed point candidates increases with increasing $N$.}
\end{center}
\end{figure}

\subsection{Identifying critical couplings}

\begin{figure}[t]
\centering
\begin{center}
\includegraphics[width=.6\hsize]{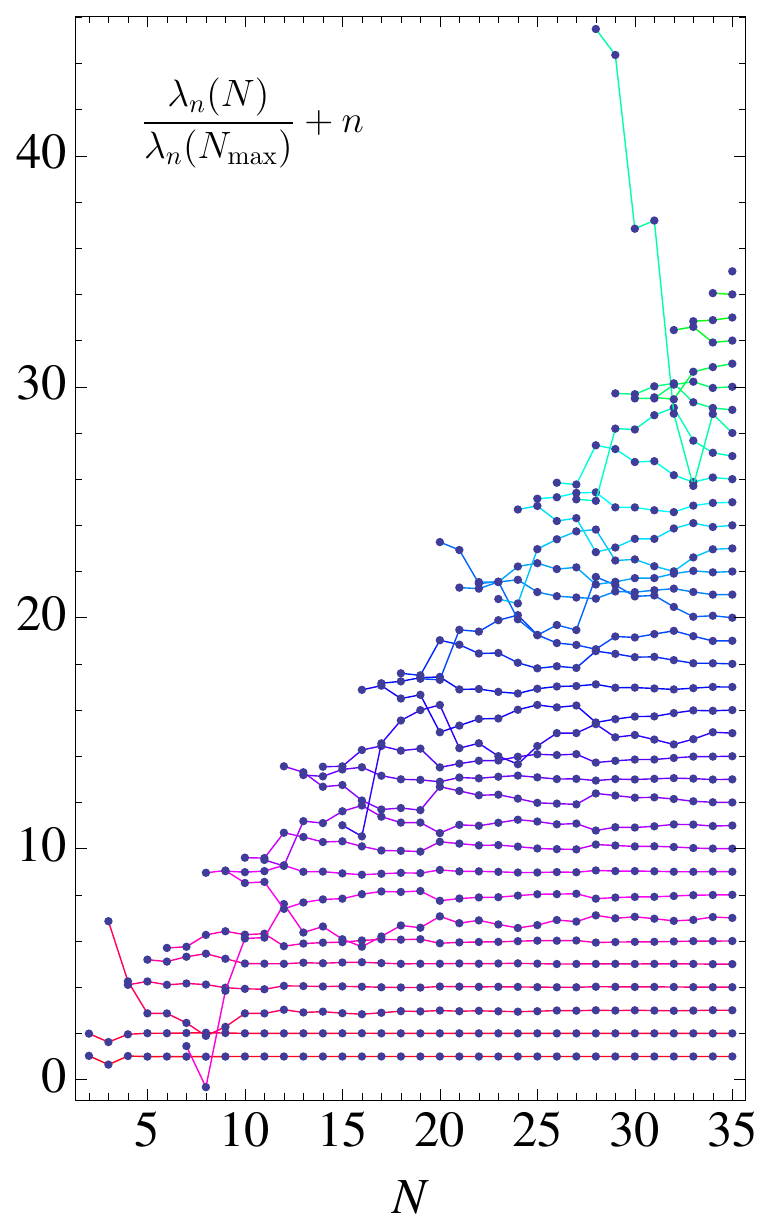}
\caption{\label{AllLambdas}Convergence pattern for all fixed point coordinates $\lambda_n(N)$ with increasing order of the polynomial approximation $N$, normalised to the values for $N_{\rm max}=35$, and with $n=0, 1,2,\cdots,34$ from bottom to top. Note that the lines for each $\lambda_n$ are shifted from each other by $n$ for better display. } 
\end{center}
\end{figure}

Following our strategy, in a first step we have obtained explicit algebraic expressions for the couplings \eq{algebraic} as functions of two free parameters $\lambda_0$ and $\lambda_1$ to high order with the help of two independent codes using {\tt Mathematica}(TM) and {\tt C++} software.   In a second step, we then need to find  the coordinates of ultraviolet fixed point(s) up to some maximal order in the polynomial expansion, $N=N_{\rm max}$ by identifying the stable roots \eq{root} for each and every order in the approximation, under the auxiliary condition \eq{boundary}. Specifically, we have used the following strategy. The algebraic expressions \eq{algebraic} are rational functions of $\lambda_0$ and $\lambda_1$, and we write them as ratios of polynomials $P_n(\lambda_0,\lambda_1)$ and $Q_n(\lambda_0,\lambda_1)$,
\begin{equation}\label{PQ}
 \lambda_n=P_n/Q_n\,.
 \end{equation}
The order of these polynomials grows rapidly with $n$. For $n=2$, $P_2$ is a cubic in $\lambda_0$ and  quadratic in $\lambda_1$, see \eq{lambda2}, whereas for $n=35$,  the polynom $P_{35}$ is of degree 264 in $\lambda_0$ and of degree  167 in $\lambda_1$, containing in total about $45\,000$ distinct terms. 
To identify stable roots at order $N$ from the boundary condition \eq{boundary}, we analyse the solutions of 
\begin{equation}\label{P}
\begin{array}{rcl}
P_{N}(\lambda_0,\lambda_1)&=&0\\[1ex]
P_{N+1}(\lambda_0,\lambda_1)&=&0\,.
\end{array}
\end{equation}  
Solutions to each of \eq{P} provide us with curves in the $(\lambda_0,\lambda_1)$ plane. We refer to theses as `nullclines'. Joint zeros are the points where the nullclines \eq{P} intersect. These provide a pair of  values \eq{root}, and thus fixed point candidates.  For consistency, we also check the nullclines of the denominators 
\begin{equation}\label{Q}
\begin{array}{rcl}
Q_{N}(\lambda_0,\lambda_1)&=&0\\[1ex]
Q_{N+1}(\lambda_0,\lambda_1)&=&0\,.
\end{array}
\end{equation}  
If \eq{P} and \eq{Q} have identical solutions, more work is needed to decide whether this is a fixed point candidate or not. We require that 
\eq{Q} does not hold for solutions to \eq{P}. We then analyse all fixed point candidates one-by-one, focussing on the regime in parameter space close to where fixed points have been found at lower orders. In principle,  
the high order of the polynomials $P_n$ may result in 
a large number of potential fixed point candidates in the complex plane. In practice, we only find a small number  of real solutions at any order, and a unique one which consistently persists from order to order. Our  guiding principle for the identification of a fixed point
are as follows. We require
\begin{equation}\label{C1}
\bullet 
\begin{array}{rl}
&{\rm consistency\ condition\ I:\ fixed\ point\ coordinates\ at\ expansion\ order\ }N\ \\ 
&{\rm should\ not\ differ\ drastically\ from\ those\ at\ order\ }N-1\,.
\end{array}
\end{equation}
If we find several fixed point candidates,  we also compute their universal eigenvalues to differentiate between them. As secondary criterion, we require that
\begin{equation}\label{C2}
\bullet 
\begin{array}{rl}
&{\rm consistency\ condition\ II:\ universal\ eigenvalues\ at\ expansion\ order\ }N\\ 
&{\rm should\ not\ differ\ drastically\ from\ those\ at\ order\ }N-1\,.
\end{array}
\end{equation}
We find that this procedure converges well. It is illustrated in Fig.~\ref{Contour16} for the example of $N=16$. Here, dashed (full)  lines correspond to the nullclines of $P_{16}$ ($P_{17}$), and the thick black line corresponds to the nullcline of $Q_{16}$. In the selected patch of parameter space we find four fixed point candidates. After detailed inspection, we conclude that the lower-left fixed point is the relevant one (full dot), linked to the fixed point found at lower orders. The other fixed point candidates (open dots) are viewed as `spurious'.

We briefly comment on additional fixed point candidates besides the main one, illustrated in Fig.~\ref{Contour8+24} for approximation order $N=8$ and $N=24$. In the search of fixed points and starting at order $N=9$ we  encounter spurious fixed points. With `spurious' we refer to fixed points which either only appear in a few selected orders in the expansion and then disappear, or whose coordinates or universal properties change drastically  from order to order, such as a change in the number of negative eigenvalues.  The arrow in Fig.~\ref{Contour8+24} (left panel) shows an example where the nullclines $Q_8, P_8$ and $P_9$ have a joint simple zero, implying that the joint zero of the nullclines $P_8=0$ and $P_9=0$ does not correspond to a fixed point.  Furthermore, with increasing $N$, the number of fixed point candidates increases, see Fig.~\ref{Contour8+24} (right panel). We conclude that the  spurious UV fixed points are artefacts of the polynomial expansion and we do not proceed their investigation any further. The physically relevant fixed point appears as an `accumulation point', surrounded by a slowly  increasing number of spurious fixed point candidates. This pattern is  similar to the one observed in simpler models at criticality, eg.~$O(N)$-symmetric scalar field theories.

\subsection{Fixed point couplings and convergence pattern}

\begin{figure}[t]
\centering
\begin{center}
\includegraphics[width=.6\hsize]{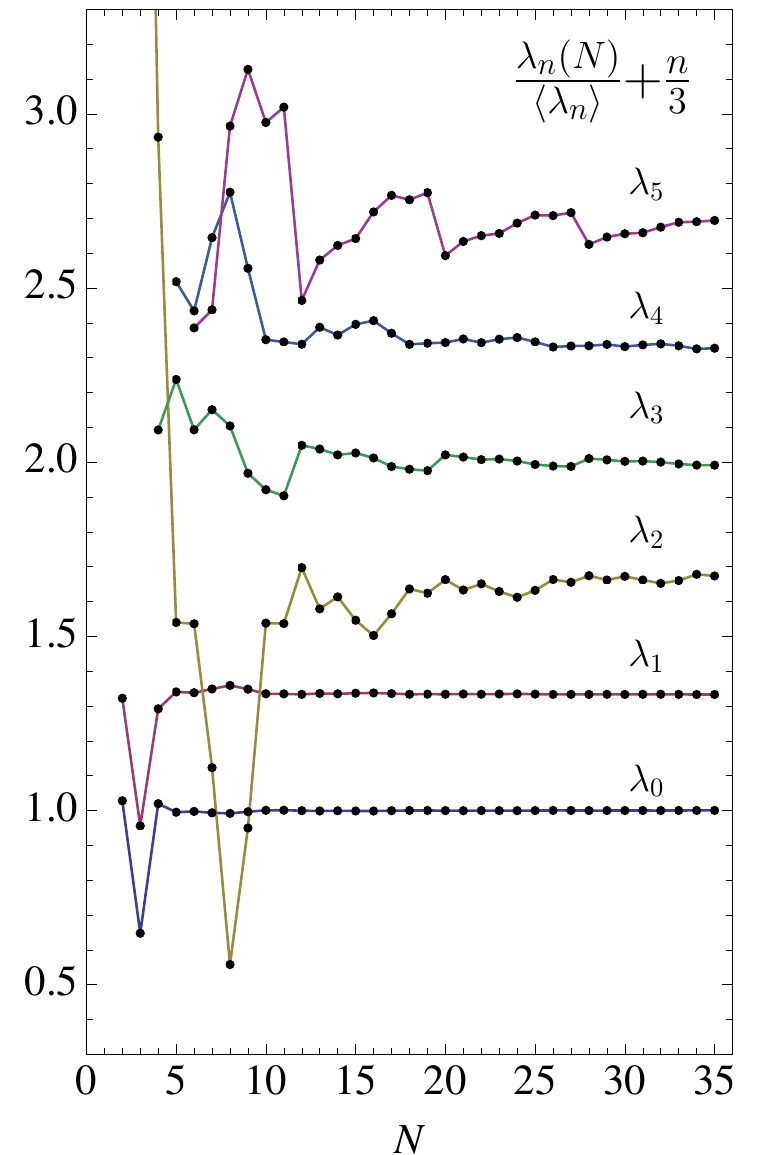}
\end{center}
\caption{\label{AllFPs}Zoom into the convergence of the first six polynomial fixed point couplings 
$\lambda_n$ 
with increasing order of the expansion $N$,  \eq{expansion0}. The couplings fluctuate about the asymptotic value $\langle \lambda_{n}\rangle$ \eq{average}, \eq{lambda*}
with a decreasing amplitude and an approximate eight-fold periodicity. Note that the convergence of the $R^2$-coupling is slower than some of the higher-order couplings. The shift term $\frac{n}{3}$ has been added for better display.} 
\end{figure}

\begin{figure}[t]
\centering
\begin{center}
\includegraphics[width=.8\hsize]{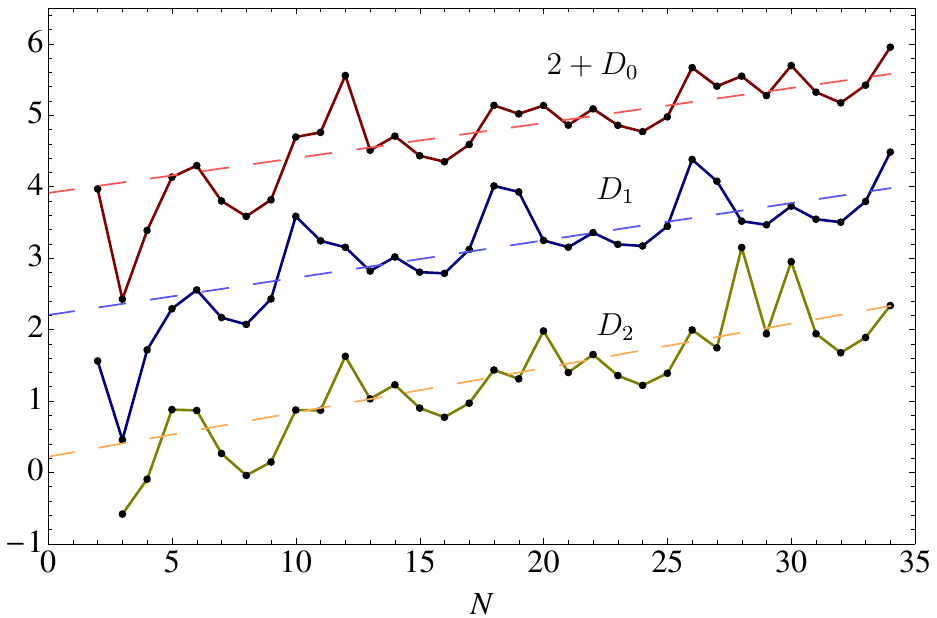}
\caption{\label{ConvergenceCheck}The rate of convergence of the three leading couplings $\lambda_0(N), \lambda_1(N)$ and $\lambda_2(N)$  towards their asymptotic values ($N\to \infty$) as given by the number of relevant digits $D_n$ \eq{digits} (from top to bottom).  The mean slopes range between $0.04-0.06$  (dashed lines), and the data points are connected through lines to guide the eye. The curve for $\lambda_0$ is shifted upwards by two units for better display.}
\end{center}
\end{figure}

Our  numerical results for the stable root \eq{root}, and thus all couplings \eq{algebraic} up to the order $N_{\rm max}=35$, are summarized in Figs.~\ref{AllLambdas},~\ref{AllFPs}, and~\ref{ConvergenceCheck}. The couplings are mostly of order one, and their signs follow, approximately, an eight-fold periodicity in the pattern
\beq\label{periodicity}
(++++----)\,.
\eeq
 Four consecutive couplings $\lambda_{3+4i}-\lambda_{6+4i}$ come out negative (positive) for odd (even) integer $i\ge 0$. Periodicity patterns such as this one often arise due to convergence-limiting singularities of the fixed point solution $f_*(\R)$ in the complexified $\R$-plane, away from the real axis. This is well-known from scalar theories at criticality where  $2n$-fold periodicities are encountered regularly \cite{Litim:2002cf,Litim:2003kf}.

\begin{center}
\addtolength{\tabcolsep}{2pt}
\begin{table*}[ht]
\setlength{\extrarowheight}{.5pt}
\normalsize
\footnotesize
\begin{tabular}{| c || c | c | c | c || c | c | c | c | } \hline
$N$ & $g_*$ & $\lambda_*$ & $g_*\times \lambda_*$  & $10\times \lambda_2$ & $\theta'$ & $\theta''$ & $\theta_2$  & $\theta_3$  \\
\hline
\hline
2 & 0.98417 & 0.12927 & 0.12722 & $$  & 2.3824 & 2.1682 & $$  & $$ \\
3 & 1.5633 & 0.12936 & 0.20222 & 0.7612 & 1.3765 & 2.3250 & 26.862 & $$ \\
4 & 1.0152 & 0.13227 & 0.13429 & 0.3528 & 2.7108 & 2.2747 & 2.0684 & $-$4.2313\\
5 & 0.96644 & 0.12289 & 0.11876 & 0.1359 & 2.8643 & 2.4463 & 1.5462 & $-$3.9106\\
6 & 0.96864 & 0.12346 & 0.11959 & 0.1353 & 2.5267 & 2.6884 & 1.7830 & $-$4.3594\\
7 & 0.95832 & 0.12165 & 0.11658 & 0.07105 & 2.4139 & 2.4184 & 1.5003 & $-$4.1063\\
8 & 0.94876 & 0.12023 & 0.11407 & $-$0.01693 & 2.5070 & 2.4354 & 1.2387 & $-$3.9674\\
9 & 0.95887 & 0.12210 & 0.11707 & 0.04406 & 2.4071 & 2.5448 & 1.3975 & $-$4.1673\\
10 & 0.97160 & 0.12421 & 0.12069 & 0.1356 & 2.1792 & 2.1981 & 1.5558 & $-$3.9338\\
11 & 0.97187 & 0.12429 & 0.12079 & 0.1354 & 2.4818 & 2.1913 & 1.3053 & $-$3.5750\\
12 & 0.97329 & 0.12431 & 0.12099 & 0.1604 & 2.5684 & 2.4183 & 1.6224 & $-$4.0050\\
13 & 0.97056 & 0.12386 & 0.12021 & 0.1420 & 2.6062 & 2.4614 & 1.5823 & $-$4.0163\\
14 & 0.97165 & 0.12407 & 0.12055 & 0.1474 & 2.4482 & 2.4970 & 1.6699 & $-$4.0770\\
15 & 0.96998 & 0.12378 & 0.12006 & 0.1369 & 2.4751 & 2.3844 & 1.5618 & $-$3.9733\\
16 & 0.96921 & 0.12367 & 0.11987 & 0.1301 & 2.5234 & 2.4051 & 1.5269 & $-$3.9590\\
17 & 0.97106 & 0.12402 & 0.12043 & 0.1398 & 2.5030 & 2.4582 & 1.5811 & $-$4.0154\\
18 & 0.97285 & 0.12433 & 0.12096 & 0.1509 & 2.3736 & 2.3706 & 1.6051 & $-$3.9487\\
19 & 0.97263 & 0.12430 & 0.12090 & 0.1490 & 2.4952 & 2.3323 & 1.5266 & $-$3.8741\\
20 & 0.97285 & 0.12427 & 0.12090 & 0.1551 & 2.5415 & 2.4093 & 1.6038 & $-$3.9805\\
21 & 0.97222 & 0.12417 & 0.12073 & 0.1504 & 2.5646 & 2.4370 & 1.5965 & $-$3.9938\\
22 & 0.97277 & 0.12428 & 0.12089 & 0.1532 & 2.4772 & 2.4653 & 1.6506 & $-$4.0332\\
23 & 0.97222 & 0.12418 & 0.12073 & 0.1498 & 2.4916 & 2.3853 & 1.5876 & $-$3.9629\\
24 & 0.97191 & 0.12414 & 0.12065 & 0.1472 & 2.5271 & 2.3999 & 1.5711 & $-$3.9596\\
25 & 0.97254 & 0.12426 & 0.12084 & 0.1503 & 2.5222 & 2.4334 & 1.5977 & $-$3.9908\\
26 & 0.97335 & 0.12440 & 0.12109 & 0.1551 & 2.4328 & 2.4025 & 1.6237 & $-$3.9734\\
27 & 0.97318 & 0.12437 & 0.12104 & 0.1539 & 2.5021 & 2.3587 & 1.5673 & $-$3.9182\\
28 & 0.97329 & 0.12436 & 0.12104 & 0.1568 & 2.5370 & 2.4047 & 1.6050 & $-$3.9728\\
29 & 0.97305 & 0.12432 & 0.12097 & 0.1549 & 2.5537 & 2.4262 & 1.6044 & $-$3.9849\\
30 & 0.97337 & 0.12438 & 0.12107 & 0.1565 & 2.4951 & 2.4527 & 1.6446 & $-$4.0165\\
31 & 0.97310 & 0.12434 & 0.12099 & 0.1549 & 2.4997 & 2.3865 & 1.5995 & $-$3.9614\\
32 & 0.97291 & 0.12431 & 0.12094 & 0.1534 & 2.5294 & 2.3980 & 1.5882 & $-$3.9606\\
33 & 0.97319 & 0.12437 & 0.12103 & 0.1547 & 2.5306 & 2.4228 & 1.6042 & $-$3.9819\\
34 & 0.97367 & 0.12445 & 0.12117 & 0.1574 & 2.4660 & 2.4183 & 1.6311 & $-$3.9846\\
35 & 0.97356 & 0.12443 & 0.12114 & 0.1567 & 2.5047 & 2.3682 & 1.5853 & $-$3.9342\\ 
\hline
\hline
mean (all) & 0.98958 & 0.12444 & 0.12320 & 0.1580 & 2.4711 & 2.3996 & $2.3513$& $-$3.9915\\
\hline
\hline
mean (cycle)& 0.97327 & 0.12437 & 0.12105 & 0.1557 & 2.5145 & 2.4097 & 1.6078 & $-$3.9746  \\
st.~dev.~(\%)& 0.02668 & 0.04025 & 0.06673 & 0.89727 & 1.122 & 1.085 & 1.265 & 0.603\\
\hline
\end{tabular}  
\vskip.1cm
\normalsize
\caption{\label{converge}
The fixed point values for the dimensionless Newton coupling $g_*$, the dimensionless cosmological constant $\lambda_*$, the $R^2$ coupling $\lambda_2$, the universal product $\lambda\cdot g$, 
and the first four exponents to various orders in the expansion, including their mean values and standard deviations.}
\end{table*} 
\end{center}

Fig.~\ref{AllLambdas} shows the convergence of all couplings with increasing $N$. Broadly speaking, we note that couplings converge well after a period of stronger fluctuations initially, in particular for a few higher order couplings, but much less so for the lower order ones. We exploit the periodicity pattern to estimate the asymptotic values of couplings $\lambda_n(N\to\infty)$ from an average over an entire cycle based on  the eight highest order values in the approximation between $N_{\rm max}-7$ and $N_{\rm max}$, 
\beq\label{average}
\langle X\rangle = \frac{1}{8}\sum^{N_{\rm max}}_{N=N_{\rm max}-7}X(N)\,,
\eeq
where $X(N)$ stands for  the  $N^{\rm th}$ order approximation for the quantity $X$. 
Fig.~\ref{AllFPs} shows the first six fixed point couplings  as a function of the order $N$ in the expansion, normalised to their asymptotic value \eq{average}. The first two couplings $\lambda_0$ and $\lambda_1$ converge rapidly towards their asymptotic values, and   settle on the percent level starting from $N\approx 10$. As expected, the convergence is slower for the higher order couplings. An interesting exception is the $R^2$ coupling $\lambda_2$, which  only just starts settling within $5\%$ of  its asymptotic value at the order  $N\approx 24$  of  the expansion, and hence much later than some of the subleading couplings. Furthermore, its value even becomes negative  once, at order $N=8$, see Tab.~\ref{converge}.  The origin for this behaviour, we believe, is that the $R^2$ coupling is the sole marginal operator in the set-up, whereas all other operators have a non-trivial canonical dimension. 
On the level of the RG $\beta$-function a non-vanishing canonical mass dimension leads to a term linear in the coupling,
\beq
\partial_t \lambda_n=(2n-4)\lambda_n +{\rm quantum\ fluctuations}\,,
\eeq
where the quantum terms are at least quadratic in the couplings.   The linear term helps stabilizing  the fixed point and the convergence of the coupled system. The absence of a linear term necessitates that all quantum terms accurately cancel amongst each other. Hence, the RG flow of classically marginal interactions is much more sensitive to the precise numerical value of couplings including higher-order ones. Therefore, to establish  the existence of the fixed point in $f(R)$ quantum gravity and its stability, it becomes mandatory to extend the expansion to  high orders,  $N\gg 8$. Interestingly, the higher order couplings $\lambda_3$ and $\lambda_4$ converge more rapidly than $\lambda_2$ and settle within $5\%$ of  their asymptotic value starting at $N\approx 12$ and $16$, respectively. This also hints at the special role played by the $R^2$ interaction. Notice also that the convergence behaviour in each coupling reflects  the underlying eight-fold periodicity pattern.

\subsection{Infinite-$N$ limit}
We can use our findings and \eq{average} to obtain an estimate for the value of polynomial couplings in the $N\to\infty$ limit. Specifically, for the  fixed point coordinates,  we find the infinite-$N$ estimates
\beq\label{lambda*}
\begin{array}{rll}
\langle\lambda_0\rangle\ =&\ \ \, 0.25574&\pm\ 0.015\%\\
\langle\lambda_1\rangle\ =&-1.02747 &\pm\ 0.026\%\\
\langle\lambda_2 \rangle\ = &\ \ \,0.01557 &\pm\ 0.9\%\\
\langle\lambda_3\rangle\ =&-0.4454 &\pm\ 0.70\%\\
\langle\lambda_4\rangle\ =&-0.3668 &\pm\ 0.51\%\\
\langle\lambda_5\rangle\ =&-0.2342 &\pm\ 2.5\%
\end{array}
\eeq
for the first six couplings. Clearly, the couplings $\lambda_0$ and $\lambda_1$ show excellent convergence with an estimated error due to the polynomial approximation of the order of $10^{-3}-10^{-4}$. The accuracy in the couplings $\lambda_2,\lambda_3$ and $\lambda_4$ is below the percent level and fully acceptable for the present study. The coupling $\lambda_5$ is the first one whose accuracy level of a few percent exceeds the one set by $\lambda_2$. Notice also that the mean value over all data  differs mildly from the mean over the last cycle of eight, further supporting the stability of the result. On the other hand, had we included all data points in the error estimate, the standard deviation, in particular for $\lambda_2$ and $\lambda_5$, would grow large due to the poor fixed point values at low orders.

The results \eq{lambda*} translate straightforwardly into fixed point values for the dimensionless Newton coupling and the cosmological constant, 
\beq\label{g*la*}
\begin{array}{rll}
\langle g_*\rangle\     =           &0.97327&\pm\ 0.027\%\\ 
\langle\lambda_*\rangle\ =&0.12437&\pm\ 0.041 \%\ .
\end{array}
\eeq
Note that because $\lambda$ is given by the ratio of $\lambda_0$ and $\lambda_1$ its statistical error is essentially given by the sum of theirs. 

In Fig.~\ref{ConvergenceCheck}  we estimate the rate of convergence for the couplings with increasing order in the expansion. To that end we compute the number of relevant digits $D_n(N)$ in the coupling $\lambda_n$ achieved at order $N$ in the approximation, using the definition \cite{Litim:2002cf,Bervillier:2007rc}
\beq\label{digits}
10^{-D_n}\equiv \left|1-\frac{\lambda_n(N)}{\lambda_n(N_{\rm max})}\right| \,.
\eeq
We could have used $\langle\lambda_n\rangle$  rather than $\lambda_n(N_{\rm max})$  in \eq{digits} to estimate the asymptotic value. Quantitatively, this makes only a small difference. The estimate for the growth rate of \eq{digits} is insensitive to this choice. 

In Fig.~\ref{ConvergenceCheck} we display the number of stable digits \eq{digits} for the first three couplings. Once more the eight-fold periodicity in the convergence pattern is clearly visible. The result also confirms that the precision in the leading fixed point couplings $\lambda_0$ and $\lambda_1$ is about $10^{-3}$ to $10^{-4}$ at the highest order in the expansion, in agreement with \eq{lambda*}. The average slope ranges between $0.04 - 0.06$, meaning that the accuracy in the fixed point couplings increases steadily by roughly one decimal place for $N\to N+ 20$. 

From the results for the fixed points, we can conclude a posteriori that the boundary condition \eq{boundary} adopted for the fixed point search is viable, as it has provided us with results stable under extension to higher order. Presumably this is linked to the relative smallness of couplings at the fixed point, all of which are of order one or smaller. We come back to this aspect in Sec.~\ref{Stability}.

\subsection{Scale-invariant products of couplings}
Fixed point couplings are non-universal. Still, some universal quantities of interest are given by specific products of couplings which remain invariant under global re-scalings of the metric field
 \begin{equation}\label{ell}
 g_{\mu\nu}\to \ell\, g_{\mu\nu}\,.
 \end{equation}
 Under \eq{ell}, the couplings scale as
\begin{equation}
\lambda_n\to\ell^{4-2n}\,\lambda_n\,.
\end{equation}
The classically marginal coupling $\lambda_2$ remains invariant under the rescaling \eq{ell}. All other couplings scale non-trivially. Consequently, various products of couplings can be formed which stay invariant under \eq{ell}. 
Such invariants may serve as a measure for the relative strength of the gravitational interactions \cite{Kawai:1989yh}.  

For couplings including up to $\lambda_4R^4$, and also using \eq{average}, we may construct six independent invariants with values
\beq\label{lambdaP}
\begin{array}{rcl}
\langle\lambda_0/\lambda_1^2\rangle\ =&\ \ \, 0.2421&\pm\ 0.07\%\\
\langle\lambda_0\lambda_3^2\rangle\ =&\ \ \,0.0507 &\pm\ 1.39\%\\
\langle\lambda_1\lambda_3 \rangle\ = &\ \ \,0.4577&\pm\ 0.71\%\\
\langle\lambda_0\lambda_4\rangle\ =&-0.0937 &\pm\ 0.49\%\\
\langle\lambda_3^2/\lambda_4\rangle\ =&-0.5411 &\pm\ 0.61\%\\
\langle\lambda_1^2\lambda_4\rangle\ =&-0.3872 &\pm\ 0.56\%\,,
\end{array}
\eeq
and similarly to higher order. Note that the error, a standard deviation, is of the same order of magnitude as the error for the first few critical exponents. Amongst these invariants, an important one is
 the product of fixed point couplings $g_*\cdot \lambda_*\equiv\lambda_0/(2\lambda_1^2)$, given in the first line of \eq{lambdaP}.  When expressed in terms of the more conventional couplings $g_*$ and $\lambda_*$, we find the universal product
\beq\label{gla*}
\langle g_*\cdot\lambda_*\rangle\     
=0.12105\pm 0.07\%
\eeq
with an accuracy which is an order of magnitude better than the one in the scaling exponents. The numerical value can be interpreted as a measure for the strength of gravitational couplings \cite{Kawai:1989yh}, inasmuch as \eq{gla*} remains unchanged under rescalings \eq{ell}, unlike the fixed point values \eq{g*la*} themselves. Furthermore, we also find that 
\begin{equation}
\langle g_*\cdot\lambda_*\rangle
=\langle g_*\rangle\cdot\langle\lambda_*\rangle
\end{equation}
within the same accuracy as \eq{gla*}, see also \eq{g*la*}. Similar results are found for \eq{lambdaP}. This supports the view that the cycle-averaged values have become independent of the underlying polynomial approximation.

\begin{figure}[t]
\centering
\begin{center}
\includegraphics[width=.6\hsize]{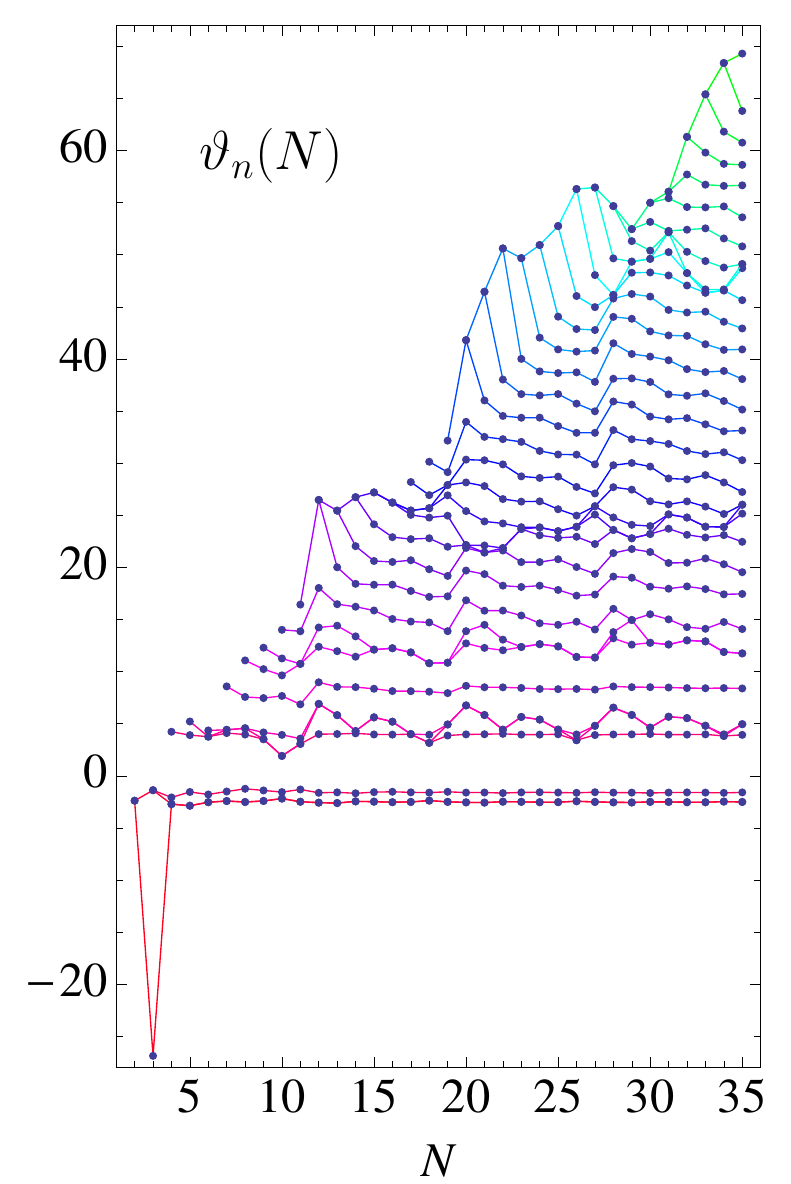}
\caption{\label{AllEigenvalues}The eigenvalues $\vartheta_n(N)$ for all approximation orders $N$ (real part if complex), sorted by magnitude. To guide the eye, lines connect the $n^{\rm th}$ smallest eigenvalue per approximation order $N$ coresponding to the columns \eq{Cn}, from bottom to top: $n=0,\cdots,N-1$. Note the eight-fold periodicity pattern in the convergence with increasing $N$, and the large negative eigenvalue which is an artefact of the $N=3$ approximation.} 
\end{center}
\end{figure}

\begin{figure}[t]
\centering
\begin{center}
\includegraphics[width=.8\hsize]{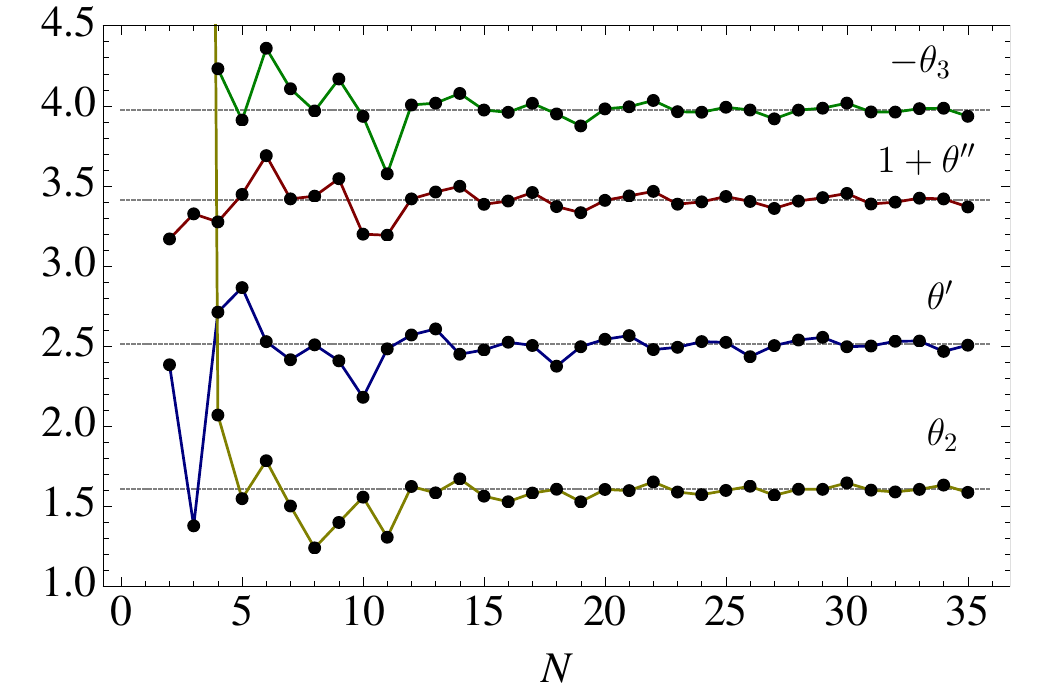}
\caption{\label{Eigenvalues}Close-up of Fig.~\ref{AllEigenvalues} into the convergence of the first four exponents  $\theta'=-{\rm Re}\, \vartheta_0$, $\theta''={\rm Im} \,\vartheta_0$, $\theta_2=-\vartheta_2$ and $\theta_3=-\vartheta_3$ -- see \eq{thetas}, \eq{thetas01}. Shown are $\theta'$ (blue line), $1+\theta''$ (red line), $\theta_2$ (yellow line)  and $-\theta_3$ (green line) together with their mean values (straight gray line). Data points converge with an eight-fold periodicity and decreasing amplitude.} 
\end{center}
\end{figure} 

\section{Scaling exponents}\label{SE}

In this section, we address universal aspects of our results
as well as the stability of the search strategy.

\subsection{Eigenperturbations and stability matrix}\label{eigenperturbations}
In critical phenomena, fixed point coordinates are often  non-universal and not measurable in any experiment. Instead, the scaling of couplings in the vicinity of a fixed point are universal. To linear order, small perturbations $\delta\,f$  from the fixed point function $f$ evolve according to 
\beq\label{delta}
 \Big(1-E_2[f]\Big)\,\partial_t\,\delta f
=
\Big(2R\,\partial_R-4+E_3[f]\Big) 
\,\delta f
\eeq
where higher order terms in $|\delta f|\ll 1$ have been supressed. Here, $E_2$ ($E_3$) are second (third) order differential operators in the dimensionless Ricci scalar $\R$. Their explicit expressions are given in \eq{E2} and \eq{E3}. Eigenperturbations $\delta f_\vartheta$ with eigenvalue $\vartheta$ obey 
\begin{equation}
\partial_t\, \delta f_\vartheta= \vartheta\cdot\delta f_\vartheta\,.
\end{equation} 
Then \eq{delta} can be used to determine the set of well-defined (finite, no poles) eigenperturbations as well as the set of  eigenvalues $\vartheta$. The sign of eigenvalues control whether eigenperturbations are relevant, marginal, or irrelevant. Notice that the fixed point solution $f$ is an integral part of \eq{delta}. 

The structure of the non-linear eigenvalue problem as given by \eq{delta} is reminiscent of the well-known Wilson-Fisher  fixed point in lower-dimensional statistical field theory. There, powerful methods have been established to reliably deduce the eigenvalues from \eq{delta}. In the polynomial approximation adopted here, the running of small deviations from the fixed point \eq{delta} can be written in terms of small deviations from the fixed point in a polynomial coordinate basis for the function $f$, leading to
\beq
\partial_t\, \delta\lambda_i=M_{ij}\,\delta\lambda_j+{\rm subleading}\,,
\eeq
where the subleading terms are higher order in $\delta\lambda=\lambda-\lambda_*$. The universal exponents then follow as the eigenvalues of the stability matrix,
\beq\label{M}
M_{ij}=\left.\frac{\partial \beta_i}{\partial\lambda_j}\right|_{\lambda=\lambda_*}
\eeq
which is, to order $N$ in the approximation, a square, real, and in general non-symmetric $N\times N$ matrix, and  $\beta_i\equiv \partial_t\lambda_i$. 
The computation of the stability matrix \eq{M} and its eigenvalues is more involved than finding the fixed points, because additional flow terms proportional to $I_1$ and $I_2$ in \eq{I} have to be taken into account as well, see \eq{I1} and \eq{I2} for explicit expressions. This is mirrored in \eq{delta} due to the presence of the differential operator $E_2$ on the LHS. Using \eq{df} and \eq{expansion0}, the $\beta$-functions can be expressed as
 \begin{equation}\label{betai}
 \beta_i=U_i+V_{ij}\beta_j
 \end{equation}
 where both $U_i$ and the matrix $V_{ij}$ are explicit functions of all couplings $\lambda_n$. One then finds
the fully resolved $\beta$-functions as 
\begin{equation}\label{betas}
 \beta=(I-V)^{-1}U
 \end{equation}
where we have suppressed indices, and $I$ denotes the identity matrix. With increasing approximation order $N$, inverting the non-numerical matrix $(I-V)$ to  find the functions $\beta_i$, and to then compute \eq{M}, becomes algebraically very demanding. Therefore, we adopt a different path and use \eq{betai} to compute the stability matrix \eq{M}  directly at the fixed point. We find
 \begin{equation}\label{MM}
 M=\left.(I-V)^{-1}\,\frac{\partial U}{\partial\lambda}\right|_{\lambda=\lambda_*}\,.
\end{equation} 
Here, the matrix $({\partial U}/{\partial\lambda})_{ij}$ stands for $\partial U_i/\partial\lambda_j$. At the fixed point, the numerical matrix $(I-V)|_*$ can be inverted reliably using standard  methods. More generally, this technique  is useful whenever the RHS of the flow contains terms proportional to the flow itself.

We have computed \eq{MM} and its sets of eigenvalues $\{\vartheta_n\}$ for all $N$ up to $N_{\rm max}=35$.  Our results are summarised in Figs.~\ref{AllEigenvalues}, \ref{Eigenvalues},~\ref{Convergence} and Tab.~\ref{converge}.  We also confirm earlier findings up  to order $N=8$, which have been obtained by first resolving \eq{betai} for $\beta_i$, and then computing \eq{M}. A discussion of the large-order behaviour of eigenvalues is deferred  to Sec.~\ref{PC}.

\subsection{Eigenvalues and scaling exponents}

We now discuss our results for the eigenvalues in more detail.  While some of them may come out as complex conjugate pairs, it is the real part of eigenvalues which decides whether the corresponding eigenperturbation is relevant, marginal, or irrelevant. Therefore at each approximation order $N$, we order the corresponding set of eigenvalues $\{\vartheta_n\}$ according to the size of their real parts, 
\begin{equation}\label{order}
{\rm Re}\,\vartheta_n(N)\le {\rm Re}\,\vartheta_{n+1}(N)\,.
\end{equation} 
We can then write these eigenvalues, for each $N$, into the rows of a matrix $T$ with elements
\begin{equation}\label{TnN}
T_{Nn}:=\vartheta_n(N)\,.
\end{equation}
This is a $N_{\rm max}\times (N_{\rm max}-1)$ matrix,
with  $n$ ranging from $n=0$ to $n=N_{\rm max}-1$, and $N$ ranging from $N=2$ to $N=N_{\rm max}$. $T$ is not a square matrix because the lowest approximation order is $N=2$ rather than $N=1$. If $n>N-1$, we have that $T_{Nn}=\vartheta_n(N)=0$.  This makes the eigenvalue matrix $T$ in \eq{TnN} a lower triangular matrix.  
The rows, columns, and diagonals of the eigenvalue matrix \eq{TnN} encode information about the convergence and stability of the polynomial approximation. By construction, each row  $T_N$ of the  matrix of eigenvalues \eq{TnN} displays the $N$ universal eigenvalues at order $N$, sorted by magnitude of their real parts \eq{order},
\beq\label{ThetaN}
T_N:=\{\vartheta_n(N)\ | \ n=0,\cdots, N-1\}
\eeq
The approximation order $N$ has $N$ eigenvalues, and hence the $N^{\rm th}$ row generically has $N$ non-zero entries.  Each column  $C_n$ of \eq{TnN} ($n$ fixed)  shows how the $n^{\rm th}$ largest eigenvalue depends on the approximation order $N$,
\beq\label{Cn}
C_n:=\{\vartheta_n(N)\ | \ N=1,\cdots, N_{\rm max}\}\,.
\eeq
Each column $C_n$ has $N_{\rm max}-n$ non-vanishing entries. Finally, we will also be interested in the diagonals of \eq{TnN},
\begin{equation}\label{Di}
D_i:=\{\vartheta_{N-i}(N)|N=\delta_{1,i}+i,\cdots,N_{\rm max}\}\,.
\end{equation}
Each diagonal $D_i$ shows the set of $i^{\rm th}$ largest eigenvalue at approximation order from $N=N_{\rm max}$ down to $N=i+\delta_{1,i}$, and has $N_{\rm max}+1-i-\delta_{1,i}$ entries. The significance of  the diagonals \eq{Di} will be discussed in Sect.~\ref{PC} in more detail.

In Fig.~\ref{AllEigenvalues}, we display the real parts of all eigenvalues \eq{ThetaN} for all approximation orders $N\le N_{\rm max}$, corresponding to the columns \eq{Cn} of the eigenvalue matrix \eq{TnN}. Each line connects the $n^{\rm th}$ largest eigenvalue from each of the sets \eq{ThetaN}, corresponding to the columns of \eq{TnN}. If the eigenvalue is a complex conjugate pair, it corresponds to a single point in Fig.~\ref{AllEigenvalues}. We note that the scaling exponents also show an eight-fold periodicity pattern in their convergence. 

Sometimes it is customary to discuss universality in terms of the critical scaling exponents $\theta_n$, to which the eigenvalues $\vartheta_n$ relate  as 
\beq\label{thetas}
\theta_n\equiv -\vartheta_n\,.
\eeq 
The results for the first few exponents \eq{thetas} are displayed in Fig.~\ref{Eigenvalues} (see Tab.~\ref{compare} for numerical values). The leading exponents are a complex conjugate pair $\theta_0=(\theta_1)^*$, and we write it as
\beq\label{thetas01}
\theta_{0,1}=\theta'\pm i\theta''\,.
\eeq 
Only the first three exponents $\theta_0,\theta_1$ and $\theta_2$ have a positive real part, whereas all other have a negative real part. 
From Fig.~\ref{Eigenvalues} we notice that the exponents  oscillate about their asymptotic values with an eight-fold periodicity and a decreasing amplitude. We estimate their asymptotic values from an average over an entire period \eq{average}, leading to the exponents
\beq\label{theta*}
\begin{array}{rll}
\langle\theta'\rangle\ =&\ \ \,2.51&\pm\ 1.2\%\\
\langle\theta''\rangle\ =&\ \ \,2.41 &\pm\ 1.1\%\\
\langle\theta_2 \rangle\ =&\ \ \, 1.61 &\pm\ 1.3\%\\
\langle\theta_3 \rangle\ =& -3.97 &\pm\ 0.6\%\,.
\end{array}
\eeq
Here, the accuracy in the result has reached the percent level for the first two real and the first pair of complex conjugate eigenvalues. The error estimate \eq{theta*} allows us to conclude  that the ultraviolet fixed point has three relevant directions. The asymptotic estimates  $\langle\theta'\rangle$, $\langle\theta''\rangle$ and $\langle\theta_3\rangle$ depend only mildly on whether the average is taken over all approximations, or only the highest ones,  see Tab.~\ref{converge}. An exception to this is the exponent $\theta_2$. The slow convergence of the fixed point coupling $\lambda_2$ has led to a very large eigenvalue at approximation order $N=3$.
Although the eigenvalue rapidly decreases by a factor of nearly $20$ with increasing $N$, its presence 
 is responsible for the overall mean value to deviate by $40\%$ from $\langle\theta_2\rangle$, \eq{theta*}, see Tab.~\ref{converge}. We therefore conclude that the large eigenvalue $\theta_2(N=3)$ is  unreliable and an artefact of the approximation $N=3$. We come back to this aspect in Sec.~\ref{Stability}.

\begin{center}
\addtolength{\tabcolsep}{4pt}
\begin{table*}[ht]
\setlength{\extrarowheight}{.5pt}
\normalsize
\footnotesize
\begin{tabular}{|c||c||r|r|r|r|r|r |} \hline

$\vartheta_{n}(N)$&
&\multicolumn{5}{c}{${}\quad\quad\quad\ \ \ $asymptotically safe fixed point}&  \\\hline
eigenvalues&
Gaussian
& $N=35$ & 31 & 23  & 15 & 11&7  \\
\hline
\hline
$\vartheta_{0}$ & $-$4 & $-$2.5047 & $-$2.4997 & $-$2.4916 & $-$2.4751 & $-$2.4818 & $-$2.4139\\ 
$\vartheta_{1}$ & $-$2 & $-$2.5047 & $-$2.4997 & $-$2.4916 & $-$2.4751 & $-$2.4818 & $-$2.4139\\ 
$\vartheta_{2}$ &\ \, 0 & $-$1.5853 & $-$1.5995 & $-$1.5876 & $-$1.5618 & $-$1.3053 & $-$1.5003\\ 
$\vartheta_{3}$ &\ \, 2 & 3.9342 & 3.9614 & 3.9629 & 3.9733 & 3.0677 & 4.1063\\ 
$\vartheta_{4}$ &\ \, 4 & 4.9587 & 5.6742 & 5.6517 & 5.6176 & 3.0677 & 4.4184\\ 
$\vartheta_{5}$ &\ \, 6 & 4.9587 & 5.6742 & 5.6517 & 5.6176 & 3.5750 & 4.4184\\ 
$\vartheta_{6}$ &\ \, 8 & 8.3881 & 8.4783 & 8.4347 & 8.3587 & 6.8647 & 8.5827\\ 
$\vartheta_{7}$ &\,10 & 11.752 & 12.605 & 12.366 & 12.114 & 10.745 & $ $\\ 
$\vartheta_{8}$ &\,12 & 11.752 & 12.605 & 12.366 & 12.114 & 10.745 & $ $\\ 
$\vartheta_{9}$ & \,14 & 14.089 & 15.014 & 15.384 & 15.867 & 13.874 & $ $\\ 
$\vartheta_{10}$ & \,16 & 17.456 & 17.959 & 18.127 & 18.336 & 16.434 & $ $\\ 
$\vartheta_{11}$ & \,18 & 19.540 & 20.428 & 20.510 & 20.616 & $ $ & $ $\\ 
$\vartheta_{12}$ & \,20 & 22.457 & 23.713 & 23.686 & 24.137 & $ $ & $ $\\ 
$\vartheta_{13}$ & \,22 & 25.158 & 25.087 & 23.686 & 27.196 & $ $ & $ $\\ 
$\vartheta_{14}$ & \,24 & 26.014 & 25.087 & 23.862 & 27.196 & $ $ & $ $\\ 
$\vartheta_{15}$ & \,26 & 26.014 & 26.048 & 26.311 & $ $ & $ $ & $ $\\ 
$\vartheta_{16}$ & \,28 & 27.235 & 28.534 & 28.734 & $ $ & $ $ & $ $\\ 
$\vartheta_{17}$ & \,30 & 30.289 & 31.848 & 32.045 & $ $ & $ $ & $ $\\ 
$\vartheta_{18}$ & \,32 & 33.131 & 34.205 & 34.361 & $ $ & $ $ & $ $\\ 
$\vartheta_{19}$ & \,34 & 35.145 & 36.606 & 36.629 & $ $ & $ $ & $ $\\ 
$\vartheta_{20}$ & \,36 & 38.069 & 39.876 & 40.008 & $ $ & $ $ & $ $\\ 
$\vartheta_{21}$ & \,38 & 40.914 & 42.258 & 49.675 & $ $ & $ $ & $ $\\ 
$\vartheta_{22}$ & \,40 & 42.928 & 44.707 & 49.675 & $ $ & $ $ & $ $\\ 
$\vartheta_{23}$ & \,42 & 45.640 & 48.011 & $ $ & $ $ & $ $ & $ $\\ 
$\vartheta_{24}$ & \,44 & 48.708 & 50.248 & $ $ & $ $ & $ $ & $ $\\ 
$\vartheta_{25}$ & \,46 & 49.101 & 52.159 & $ $ & $ $ & $ $ & $ $\\ 
$\vartheta_{26}$ & \,48 & 49.101 & 52.159 & $ $ & $ $ & $ $ & $ $\\ 
$\vartheta_{27}$ & \,50 & 50.800 & 52.291 & $ $ & $ $ & $ $ & $ $\\ 
$\vartheta_{28}$ & \,52 & 53.591 & 55.422 & $ $ & $ $ & $ $ & $ $\\ 
$\vartheta_{29}$ & \,54 & 56.658 & 56.048 & $ $ & $ $ & $ $ & $ $\\ 
$\vartheta_{30}$ & \,56 & 58.625 & 56.048 & $ $ & $ $ & $ $ & $ $\\ 
$\vartheta_{31}$ & \,58 & 60.755 & $ $ & $ $ & $ $ & $ $ & $ $\\ 
$\vartheta_{32}$ & \,60 & 63.796 & $ $ & $ $ & $ $ & $ $ & $ $\\ 
$\vartheta_{33}$ & \,62 & 69.299 & $ $ & $ $ & $ $ & $ $ & $ $\\ 
$\vartheta_{34}$ & \,64 & 69.299 & $ $ & $ $ & $ $ & $ $ & $ $\\\hline
\end{tabular}  
\normalsize
\caption{\label{compare}The  large-order behaviour of asymptotically safe eigenvalues  for a selection of orders  $N$  in the polynomial expansion in comparison with Gaussian eigenvalues. If the eigenvalues are a complex conjugate pair,
 only the real part is given.}
\end{table*} 
\end{center}

\subsection{Gap in the eigenvalue spectrum}

At the Gaussian fixed point, the eigenvalue spectrum is equidistant, with $\theta_{G,n}=4-2n$ for $n\ge 0$, see \eq{classicalscaling}. Consequently, the model has two relevant and one marginal coupling. The least relevant eigenvalue, $\theta_{G,2}$, is marginal. We denote the distance between the least relevant and the least irrelevant eigenvalue as the `gap' $\Delta$ in the eigenvalue spectrum. The gap in the spectrum is an observable, and its value is interesting in that it captures information about quantum corrections to the borderline between relevancy and marginality or irrelevancy of eigenoperators. Classically, the gap  reads $\Delta_G\equiv \theta_{G,2}-\theta_{G,3}$, meaning
\begin{equation}\label{GapGauss}
\Delta_G=2\,.
\end{equation}
At the interacting fixed point detected here, the eigenvalues $\theta_2$ and $\theta_3$ continue to mark the divide between relevant and irrelevant couplings in the UV. The smallest relevant eigenvalue $\theta_2=1.61$ is much larger and thus more relevant than the classically marginal eigenvalue $\theta_{G,2}=0$. At the same time the most relevant of the irrelevant eigenvalues, $\theta_3=-3.97$, is less relevant than the perturbative estimate  $\theta_{G,3}=-2$. In consequence, we find that the gap $\Delta=\theta_3-\theta_2$ between the smallest relevant (in absolute size) and the smallest irrelvant eigenvalues widens due to asymptotically safe interactions,
\begin{equation}\label{gap}
\Delta_{\rm UV}\approx 5.58\,. 
\end{equation}
This is much larger than the gap at the Gaussian fixed point, $\Delta_{\rm UV}>\Delta_{\rm G}$. The enhancement of the gap should be seen as a consequence of the quantum dynamics.  Numerically, the result is stable from order to order in the approximation.    

\begin{figure}
\centering
\begin{center}
\includegraphics[width=.95\hsize]{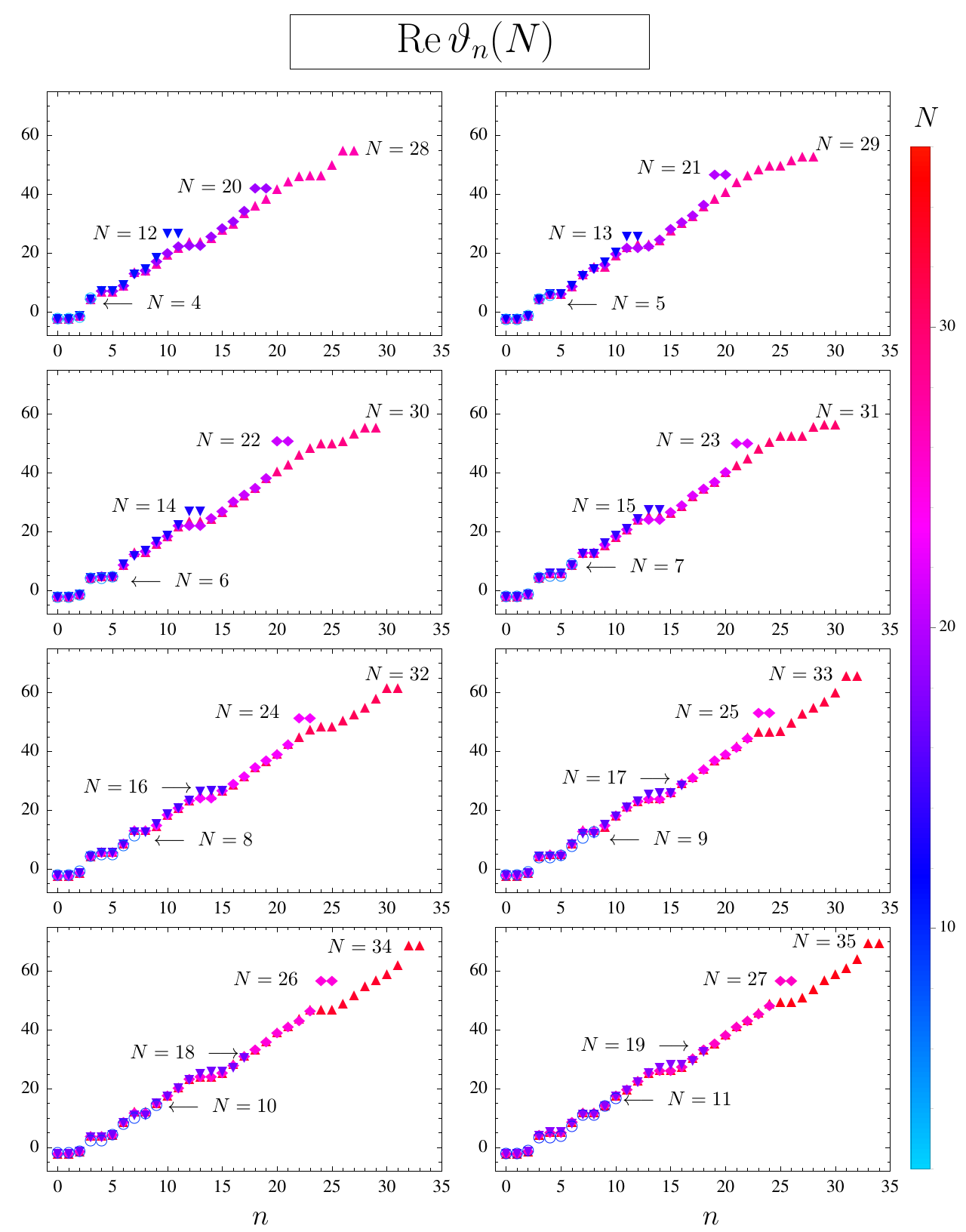}
\caption{\label{Convergence} Convergence and eight-fold periodicity pattern of the real part of eigenvalues $\vartheta_n(N)$ with increasing order of the approximation $N$, covering the range $N=4,\cdots,35$. From top left to bottom right, each sub-plot shows four sets $T_N$ of eigenvalues \eq{ThetaN} with approximation orders differing by multiples of the approximate periodicity $\Delta N=8$. In each sub-plot, different symbols and colour coding are used 
to distinguish the data sets with decreasing $N$; colour-coding as indicated.}
\end{center}
\end{figure}

\subsection{Convergence and periodicity}\label{CP}

Both the fixed point coordinates and the universal eigenvalues display an eight-fold periodicity pattern in their convergence pattern. This becomes transparent in Fig.~\ref{Convergence} which displays our results for the eigenvalues (real part if complex). To simplify the order-by-order comparison, in each of the eight sub-plots we compare four eigenvalue  sets $T_N$, whose approximation orders differ by multiples of the periodicity $\Delta N=8$. Thereby we cover results from all data sets $T_N$ between $N=4$ and $N=35$. We note that two neighboring points in Fig.~\ref{Convergence} with the same magnitude indicate a complex conjugate pair of eigenvalues.

We notice that the eigenvalues of approximation orders differing by multiples of the periodicity are essentially on top of each other, except for the highest eigenvalues. More often than not, the highest eigenvalues are a complex conjugate pair, which settle towards their physical values only once further higher order couplings are retained.
A few conclusions can be drawn from Fig.~\ref{Convergence}: Firstly, the first few eigenvalues are remarkably stable to all orders $N$, including the leading complex conjugate pair. This result is at the root for the high accuracy in the estimates \eq{theta*}. Secondly, we also notice that eigenvalues remain stable provided the approximation is extended by $\Delta N=8$, consistent with the eight-fold periodicity pattern observed in the underlying fixed point values. Thirdly, we observe that the size of exponents grows linearly with $n$, roughly as $\vartheta_n\approx 2n$ for large $n$. The largest eigenvalues at each $N$ are either a complex conjugate pair, or real. If the largest eigenvalues are a complex conjugate pair, they stick out in magnitude and deviate visibly from estimates for larger $N$ for the same exponent $\theta_n$. With increasing $N$, however, these eigenvalues rapidly decrease, and some but not all of them turn into real eigenvalues. If the largest eigenvalue is real, its size compares well with estimates from approximations with larger $N$. 

\begin{figure}[t]
\centering
\begin{center}
\includegraphics[width=.7\hsize]{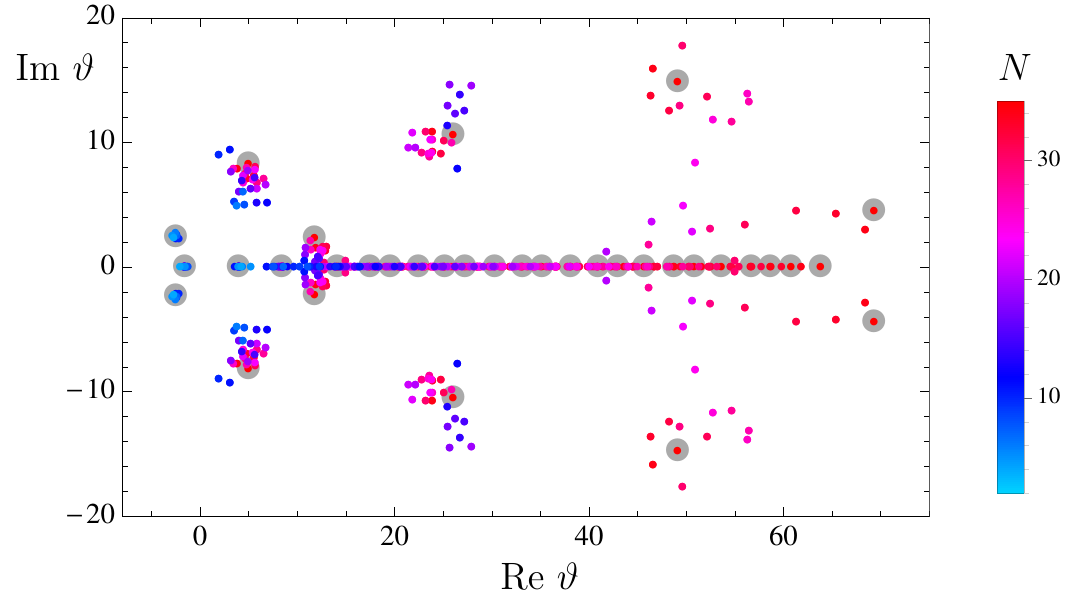}
\caption{\label{Complex} Distribution of  eigenvalues at the ultraviolet fixed point  in the complex plane. Gray-filled circles underlay the results for the eigenvalues $\vartheta_n$ at the highest approximation order $N=35$. Small coloured circles indicate eigenvalues for the approximations $4\le N\le 35$.   Most eigenvalues are real. The imaginary part of eigenvalues are more sensitive to the approximation and show slower convergence.}
\end{center}
\end{figure}

\begin{figure}[t]
\centering
\begin{center}
\includegraphics[width=.9\hsize]{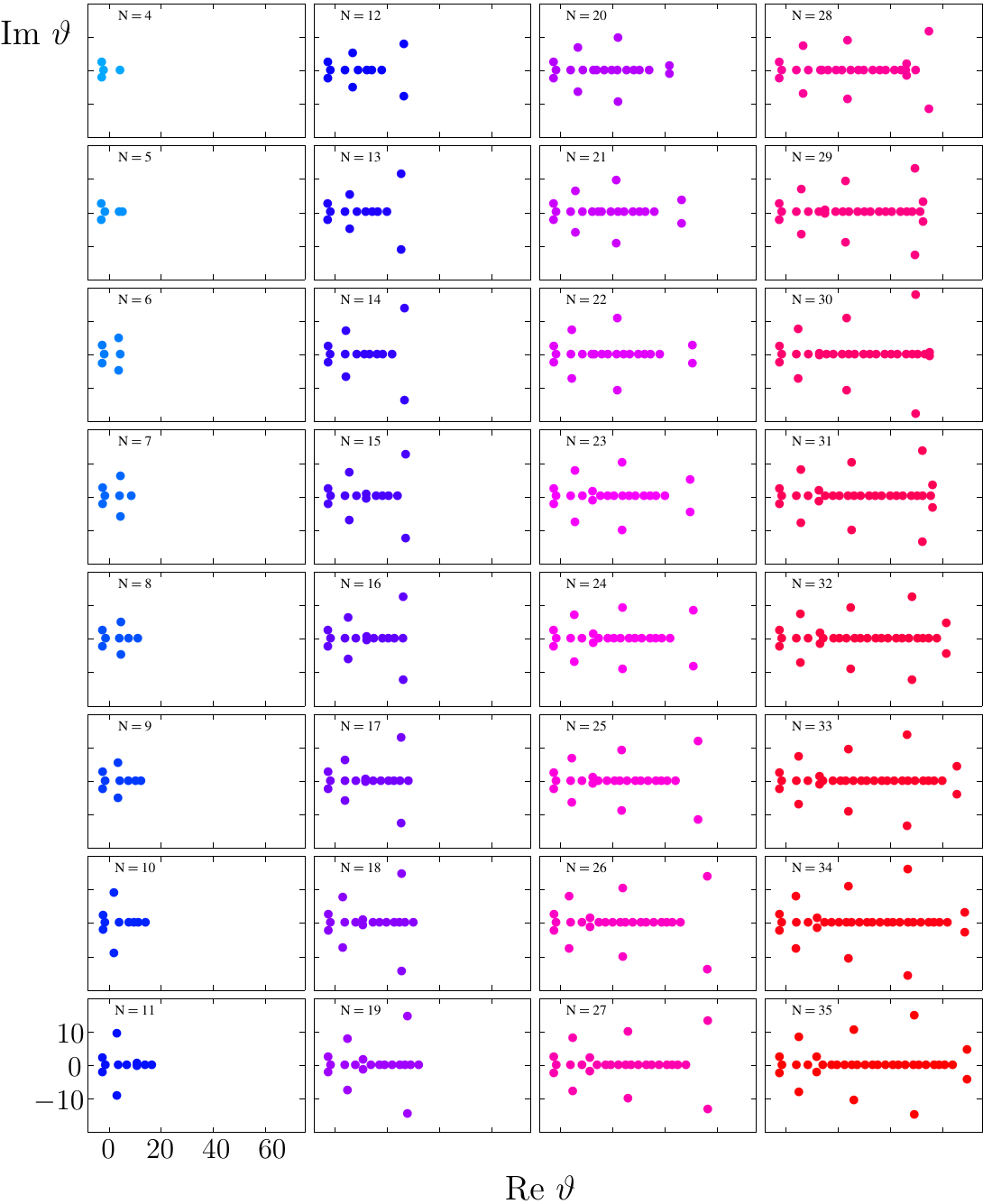}
\caption{\label{ComplexAll} Order-by-order evolution of the eigenvalue spectrum at the ultraviolet fixed point in the complex plane. Shown are 32 shots for the order-by-order convergence  of eigenvalues for all approximations from $N=4$ (top left) to $N=35$  (bottom right). Axes and colour coding exactly as in Fig.~\ref{Complex}.}
\end{center}
\end{figure}

\begin{figure}[t]
\centering
\begin{center}
\includegraphics[width=.7\hsize]{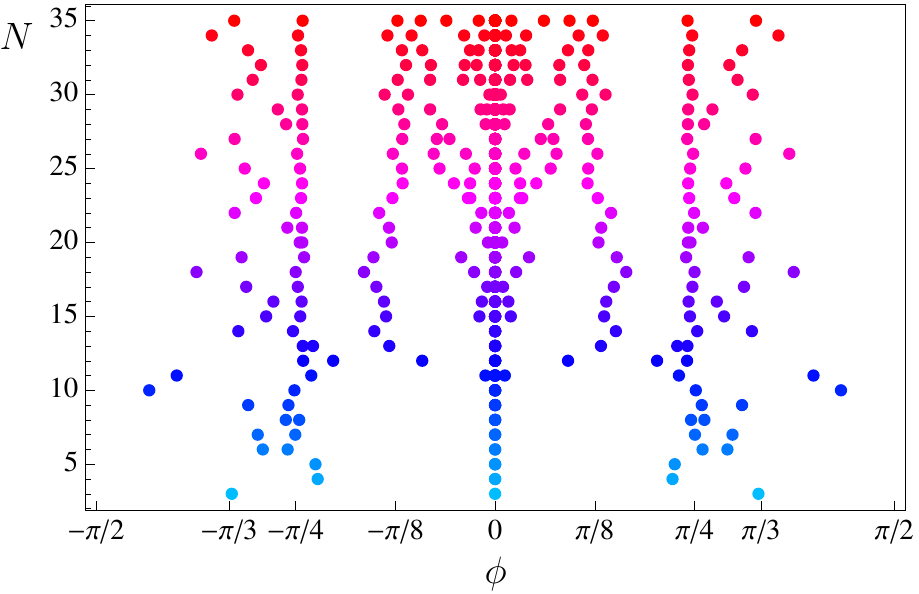}
\caption{\label{Angles} Tomography of the  angles $\phi$ \eq{angles}  of eigenvalues in the complex plane at the ultraviolet fixed point  as a function of the order of approximation $4\le N\le 35$. Most eigenvalues are real with $\phi=0$. The leading complex pair of eigenvalues settles close to $\phi\approx \pm\pi/4$. The next-to-leading and the next-to-next-to-leading complex conjugate pair appear close to the angles $\pm \pi/3$ and $\pm\pi/8$, respectively, and their convergence is slower. Colour-coding as in Figs.~\ref{Complex} and~\ref{ComplexAll}.}
\end{center}
\end{figure}

Fig.~\ref{Complex} and Fig.~\ref{ComplexAll}   show our results for all scaling exponents including their imaginary parts in all approximations considered. In Fig.~\ref{Complex}, the large gray dots indicate the results for the approximation $N=35$. Smaller coloured dots indicate the results for all other approximations $4\le N\le 34$. Most eigenvalues are real, and many eigenvalues never develop an imaginary part. Those which do show a stronger dependence on the approximation order, except for  the smallest complex conjugate pair $\vartheta_{0,1}$ which is confirmed to be remarkably stable.
The imaginary parts of the subleading pairs $\vartheta_{4,5}$ and $\vartheta_{7,8}$ have varied more strongly with the order of the approximation. For some of the higher-order eigenvalues such as the pair $\vartheta_{33,34}$, the order of our approximation is not yet good enough to settle whether these will come out real or complex in the asymptotic limit $N\to \infty$. In Fig.~\ref{ComplexAll} the convergence of scaling exponents in the complex plane is made transparent for all $4\le N\le 35$. From order to order, the small eigenvalues start converging rapidly. The higher eigenvalues are often a complex conjugate pair, and with increasing order these either settle to complex values, or bifurcate into real ones. Evidently, there are no large jumps  or discontinuous changes in the order-by-order development of the eigenvalue spectrum. 
In Fig.~\ref{Angles} we display the angles
\beq\label{angles}
\phi_n=\arctan \frac{{\rm Im}\,\vartheta_n}{{\rm Re}\,\vartheta_n}
\eeq
between real and imaginary part of all eigenvalues in \eq{ThetaN} for all approximation orders $N$. The majority of eigenvalues is real with $\phi=0$. The leading complex conjugate pair $\vartheta_{0,1}$ occurs under an angle of $\phi\approx \pm \pi/4$. The angle converges visibly fast with increasing $N$. The next-to-leading and the next-to-next-to-leading complex conjugate pairs $\vartheta_{4,5}$ and $\vartheta_{7,8}$ appear with angles close to $\pm \pi/3$ and $\pm\pi/8$, respectively. Their convergence is much slower though.
We conclude that the overall convergence of exponents is quite good. The largest eigenvalues per approximation order can probably not be trusted quantitatively if these are a complex conjugate pair.

\subsection{Interactions and degenerate scaling}\label{deg}
We briefly comment on the appearance of complex conjugate pairs of scaling exponents. The matrix $M$ in \eq{M} is in general a non-symmetric real matrix, and therefore some of its eigenvalues may become complex. At the asymptotically safe fixed point in $f(R)$ quantum gravity, we find several such complex conjugate pairs of exponents, including the pairs $\vartheta_{0,1}$, $\vartheta_{4,5}$, and $\vartheta_{7,8}$ which persist systematically even to high approximation order. 
One may wonder whether complex eigenvalues are a stable characteristic of fixed point gravity or limitations of our approximations. 

Here, we wish to point out that complex eigenvalues indicate, prima fac\ae, a degeneracy within the coupled system at criticality, which can be understood as follows. In the limit  where quantum fluctuations are absent, the matrix $M$ becomes diagonal, and its eigenvalues real
\begin{equation}
\vartheta=\vartheta^*\,.
\end{equation} 
Quantum fluctuations are responsible for the occurrence of off-diagonal entries of the matrix $M$.  If the eigenvalues remain real (and non-degenerate), then two linearly independent eigenperturbations can unambiguously be distinguished by their decay (or growth) rate with RG scale. 
On the other hand, if some of the interaction-induced off-diagonal entries happen to be numerically large, the stability matrix \eq{M} can develop complex conjugate pairs of eigenvalues 
\begin{equation}
\vartheta\neq \vartheta^*\,.
\end{equation}
As a consequence, the RG scaling of  two linearly independent eigenperturbations becomes entangled, to the extend that 
their envelope decay (or growth) rate with RG scale is governed by exactly one and the same universal index,
the real part of their eigenvalue
\begin{equation}
{\rm Re}\,\vartheta\,.
\end{equation} 
The sole difference between these eigenperturbations then relates to a relative phase, controlled by the eigenvalue's imaginary part, which thereby serves as a measure for the entanglement: the larger $|$Im$\,\vartheta |$ the larger the entanglement between eigenperturbations, and vice versa. The presence of a complex eigenvalue thus implies that the leading behaviour of the associated eigenperturbations  is exactly the same, with differences  related to phase shifts appearing at subleading level.

It is conceivable that degeneracies are lifted through additional interactions, neglected in the present approximation. In fact, adding more interaction terms can reduce large off-diagonal entries of the stability matrix into smaller ones, leading to the occurrence of real exponents within the larger system of couplings. Known examples which lift the degeneracy of $\vartheta_{0,1}$ include  Einstein-Hilbert gravity in higher dimensions \cite{Fischer:2006fz,Fischer:2006at}, the inclusion of  fourth order derivative couplings \cite{Benedetti:2009gn}, extended ghost interactions in Einstein-Hilbert gravity \cite{Christiansen:2012rx}, or the inclusion of matter fields. More work is required to decide whether complex scaling exponents survive in the physical theory, or whether they arise due to our approximations by eg.~neglecting other interaction terms.

\section{Non-perturbative boundary conditions}\label{Stability}
In this section, we test the stability of the fixed point  solution against variations of the boundary condition \eq{boundary}, and put forward the idea of self-consistent boundary conditions.

\subsection{Boundary conditions}\label{BC}
 Thus far we have identified fixed points and their eigenvalues by increasing the order of expansion one by one, achieving a coherent picture for a non-trivial UV fixed point with the help of free boundary conditions
 \beq
 \label{boundary2}
\begin{split}
\lambda_N        \, &=\ 0\\
\lambda_{N+1} \,&=\ 0
\end{split}
\eeq
for the fixed point search. The stability in the fixed point coordinates with increasing order confirms that we have identified one and the same underlying fixed point at each and every order in the expansion. 

To clarify the role of the boundary condition \eq{boundary2} we perform the fixed point search at order $N$ by 
using a one-parameter family of boundary conditions which are informed by the non-perturbative fixed point values \eq{lambda*}, namely
\beq\label{boundary*}
\begin{array}{rl}
\lambda_N         &=\alpha \cdot \lambda^{np}_{N}       \\[1ex]
\lambda_{N+1} &=\alpha\cdot \lambda^{np}_{N+1}\,.
\end{array}
\eeq
Here, the numbers $\lambda_i^{np}$ stand for the non-perturbative values of the higher order couplings which are not part of the RG dynamics at approximation order $N$. In other words, we use the asymptotic estimates
 \eq{lambda*} as input. More generally, boundary conditions such as \eq{boundary*} could be interpreted as the presence of an external non-dynamical gravitational background field without any quantum dynamics of its own. The free parameter $\alpha$ is then used to interpolate between the original `free' boundary condition \eq{boundary} $(\alpha=0)$ adopted initially to detect the fixed point, and an improved boundary condition where the choice for the higher order couplings is guided by the by-now known non-perturbative result $(\alpha=1)$ obtained from the $\alpha=0$ search. For notational simplicity, we refer to results achieved at approximation  order $N$ with  boundary condition \eq{boundary*}  as the `$N_\alpha$-approximation'.  In this convention our results in Tab.~\ref{converge} correspond to the $N\equiv N_{\alpha=0}$ approximation.

From the point of view of the RG flow, the boundary condition \eq{boundary*} with $\alpha=1$ means that we splice non-perturbative information originating from higher orders back into a smaller sub-system of relevant couplings. 
The boundary condition then acts like  a `non-perturbative background' generated from  non-dynamical higher-order couplings.  
Evidently, by virtue of the exact recursive relations amongst the fixed point couplings \eq{algebraic}, we find that the fixed point coordinates in any of the approximations $N_{\alpha=1}$ are given exactly by the asymptotic values \eq{lambda*}. Hence, the primary effect of the non-perturbative 
boundary condition is to re-align the fixed point coordinates with those achieved for asymptotically large approximation order. 

A secondary effect relates to the impact of the non-dynamical higher order couplings on the universal scaling exponents for the dynamical couplings. This can be seen as follows. At approximation order $N_{\rm max}$, the stability matrix $M$ is a $N_{\rm max} \times N_{\rm max}$ matrix. In the full theory  the model contains infinitely many couplings $N_{\rm max}\to\infty$, and the stability matrix $M$ \eq{M} would formally become infinite-dimensional.
Suppose now that we only wish to retain $N<N_{\rm max}$ couplings as dynamical ones, but that we have some information about fixed point values for the remaining non-dynamical couplings $\lambda_i$ with $N<i\le N_{\rm max}$. The full stability matrix then decomposes as
\beq\label{MABCD}
M=\left(\begin{array}{cc}
A&B\\
C&D
\end{array}
\right)
\eeq
into submatrices $A, B, C$ and $D$. Here, $A$ is the $N\times N$ sub-matrix corresponding to the $N$ retained `dynamical' couplings. The entries of the $(N_{\rm max}-N)\times N$ matrices $B$ and $C^T$ decode the mixing between the `dynamical' and the `non-dynamical' couplings. Finally, the $(N_{\rm max}-N)\times (N_{\rm max}-N)$ matrix $D$ mainly encodes the mixing of the suppressed couplings amongst themselves. At approximation order $N$, the eigenvalues of $M$ reduce to those of the matrix $A$, and the admixture due to $B, C$ and $D$ is neglected. The eigenvalues of $A$, however, are still informed by all fixed point couplings $\lambda_n$ up to $n=N+2$, including non-dynamical ones. As such, the eigenvalues of the matrix $A$ are sensitive to the boundary condition such as \eq{boundary*} imposed on the non-dynamical couplings. 

\begin{figure}[t]
\centering
\begin{center}
\includegraphics[width=.9\hsize]{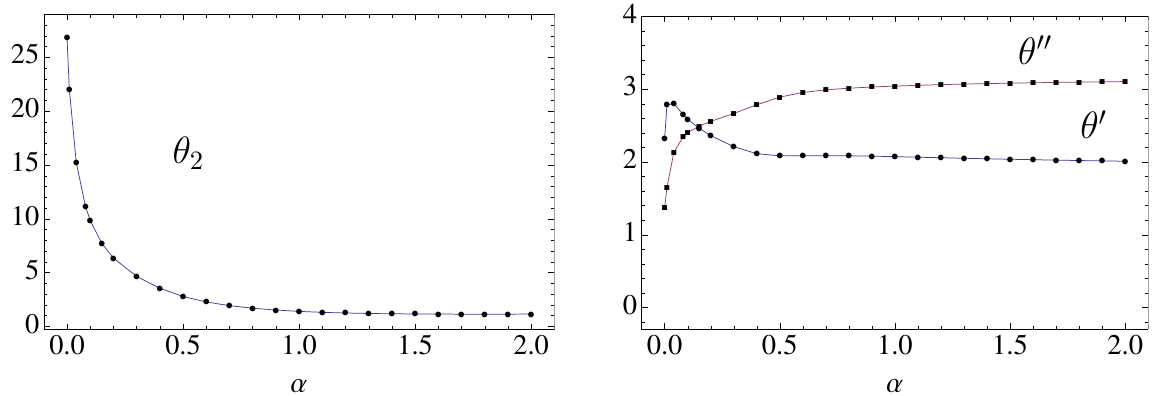}
\caption{\label{Continuity3}Continuity of the fixed point for $R^2$ gravity, shown in terms of the critical exponents $\theta_2(\alpha)$  (left panel) and $\theta'(\alpha)$,  $\theta''(\alpha)$ (right panel) 
as functions of $\alpha$. The curves smoothly interpolate between \eq{N3}
$(\alpha=0)$ and \eq{N3*} $(\alpha=1)$. Note the substantial decrease of $\theta_2$ with increasing $\alpha$. The dependence on $\alpha$ becomes very weak already around the prefered value $\alpha\approx 1$.}
\end{center}
\end{figure}

\subsection{Effects of non-dynamical higher-order couplings}

Next we analyse this effect quantitatively for the case with three and four dynamical couplings. We start with $N=3$. We recall the result in the  $N_{\alpha=0}=3$ approximation, where the exponent $\theta_2$ deviates substantially from the asymptotic value,
\beq\label{N3}
\begin{array}{rl}
\theta'&=1.3765\\
\theta''&=2.3250\\
\theta_2&= 26.862\,.
\end{array}
\eeq 
Adopting now the 
improved boundary condition as described above, we find for $N_{\alpha=1}=3$ the scaling exponents
\beq\label{N3*}
\begin{array}{rl}
\theta'&=3.0423\\
\theta''&=2.0723\\
\theta_2&=1.3893\,.
\end{array}
\eeq
The effect is substantial. Most notably, the exponent $\theta_2$ in \eq{N3*} is vastly different from its value at $N_0=3$, \eq{N3}, and all three values \eq{N3*} are now significantly closer to the asymptotic ones \eq{theta*}. Quantitatively, at order $N_0=3$ the exponents $(\theta', \theta_2)$ differ from the asymptotic ones \eq{theta*} by about $(50\%, 1700\%)$. This is reduced to  $(15\%,15\%)$ as soon as the correct background values for the non-dynamical couplings are retained, \eq{N3*}. The universal phase $\theta''$ stays within $5\%$ throughout.
The remaining difference between \eq{N3*} and \eq{theta*} is due to the fact that the RG dynamics of higher order couplings is not taken into account in the former, encoded in the matrices $B, C$ and $D$ in \eq{MABCD}. Empirically, we conclude that only about $15\%$ of the scaling exponents' values  is attributed to the dynamics of all higher order interactions. Conversely, about $85\%$ of their values is due to the dynamics of the three leading couplings, in conjunction with the correct fixed point value for non-dynamical higher order couplings.

We now turn to the next approximation order, $N=4$. The results for $N_{\alpha=0}=4$, given in Tab.~\ref{converge}, are already closer to the high-order result than  those for $N_{\alpha=0}=3$, owing to the presence of the $R^3$ interaction. Therefore, we may expect that an improved boundary condition which now affects the non-dynamical $R^4$ and $R^5$ couplings should only lead to small modifications. Quantitatively, for $N_{\alpha=1}=4$, we find
\beq\label{N4*}
\begin{array}{rl}
\theta'&=\ \ \, 2.9010\\
\theta''&=\ \ \, 2.3042\\
\theta_2&=\ \ \, 1.8336\\
\theta_3&=-2.9824\,.
\end{array}
\eeq
This should be compared with  the approximation $N_{\alpha=0}=4$ given in Tab.~\ref{converge},
and with the asymptotic values \eq{theta*}. Already at this order, the effect is less pronounced. 
    It is very encouraging that the dynamical effect of the higher-order interactions only leads to a comparatively small quantitative shift with respect to \eq{N3*}, without affecting the qualitative result. The results \eq{N3*}, \eq{N4*} also establish that the fixed point of the system is already carried reliably by a low-order approximation, provided the boundary condition is informed by the fixed point coordinates to high order. This pattern persists to higher $N$.

\begin{figure}[t]
\centering
\begin{center}
\includegraphics[width=.95\hsize]{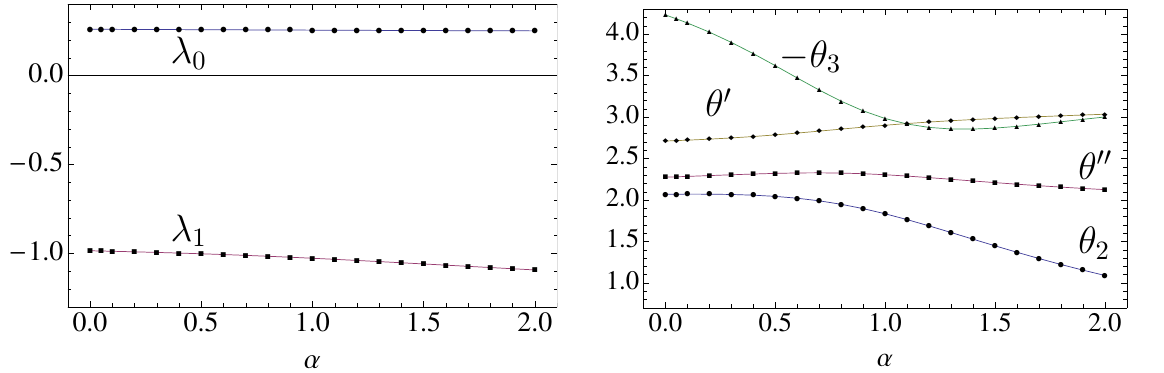}
\caption{\label{Continuity4}Continuity of the fixed point for $R^3$ gravity, showing the coordinates (left panel) and the exponents (right panel)
as functions of $\alpha$. The result smoothly interpolates between the data in Tab.~\ref{compare} $(\alpha=0)$ and \eq{N4*} $(\alpha=1)$. Note that the dependence on $\alpha$ becomes very weak already close to the preferred value $\alpha\approx 1$.}
\end{center}
\end{figure} 

\subsection{Continuity in the boundary condition}\label{Continuity}
At low order in the approximation order $N$, in particular at $N_{\alpha=0}=3$, the coordinates and scaling exponents deviate more strongly from their asymptotic value. This raises questions as to whether these solutions are spurious rather than images of the physical fixed point, and whether there are ways of improving the low-order results. To answer this question, we assess the continuity of our results subject to  the boundary condition.  
We vary $\alpha$ over some range, $0\le\alpha\le 2$ to understand how strongly the scaling exponents are affected by the boundary condition. Our results to order $N=3$ in the approximation are given in Fig.~\ref{Continuity3}. We note that all three exponents vary strongly with $\alpha$ close to the boundary condition \eq{boundary}, and up to $\alpha< 1/2$, but substantially less so once $\alpha>1/2$. Interestingly, this result also establishes that the fixed point at order $N_{\alpha=0}=3$ is in fact continuously connected with the improved result $N_{\alpha=1}=3$. Most importantly,  the relative variations with $\alpha$ are small,
\beq\label{variation}
\begin{array}{rl}
\displaystyle
\left.
\frac{\partial \ln \theta'}{\partial\ln  \alpha}
\right|_{\alpha=1}
& \approx   -0.0339\\[3ex]
\displaystyle
\left.
\frac{\partial \ln \theta''}{\partial\ln  \alpha}
\right|_{\alpha=1}
&\approx \ \ \, 0.0383\\[3ex]
\displaystyle
\left.
\frac{\partial \ln \theta_2}{\partial\ln  \alpha}
\right|_{\alpha=1}
&\approx -0.761\,.
\end{array}
\eeq
We  conclude that imposing self-consistent boundary conditions, provided they are available, improves the solution  for the low order couplings and scaling exponents.

The corresponding results for $N=4$ are shown in Fig.~\ref{Continuity4}. We note that the fixed point coordinates depend weakly on $\alpha$. In addition, the universal eigenvalues show a weak and smooth dependence on $\alpha$, and the value $\alpha=1$ is not distinguished. 
We conclude that the fixed point is stable under variations of the boundary condition imposed on the higher-order couplings. These results establish the self-consistency of the fixed point solution established here.

\subsection{Discussion}
We briefly discuss our results in the light of earlier findings \cite{Lauscher:2002sq,Codello:2008vh,FischerLitim2007,Rechenberger:2012pm}. With increasing approximation order, we have established that the perturbatively marginal $R^2$ coupling shows a much slower rate of convergence than the perturbatively relevant and some of the perturbatively irrelevant couplings. In fact, roughly $N\approx 24$ orders in the Ricci scalar are needed to ensure that the $R^2$ coupling  stays  within 5\% of its  large-$N$ estimate. The $R^0, R^1, R^3$ and $R^4$ couplings, for comparison,  achieve the same level of accuracy starting  already at the much lower orders $N=4, 4, 12$ and $16$, respectively. 
The comparatively slower convergence of the $R^2$ coupling is related to its vanishing canonical mass dimension, and  also to the underlying eight-fold periodicity pattern, highlighting again the importance of a high-order study. A side effect of this is the occurrence of a  numerically large eigenvalue $\theta_2$ in \eq{N3} at approximation order $N=3$. This has been observed in earlier studies \cite{Lauscher:2002sq,Codello:2008vh,FischerLitim2007,Rechenberger:2012pm} which have retained the same operator content (up to including $R^2$ invariants), irrespectively of the finer details of the implementation of the RG flow. This is now understood as an artefact of the boundary conditions \eq{boundary} adopted for the fixed point search.
The use of improved boundary conditions without  otherwise changing the approximation  already proves sufficient to stabilise both the fixed point coordinate and the exponents. Comparing  the improved low-order result~\eq{N3*} with the high-order results in Fig.~\ref{Eigenvalues}, we have established that the eigenvalues settle at values much closer to their $N\to\infty$ extrapolation without the necessity of introducing fully dynamical higher order invariants into the action.

\section{Bootstrap for asymptotic safety}\label{PC}

In this section, we discuss our results in the light of the asymptotic safety conjecture for gravity and a bootstrap test put forward in \cite{Falls:2013bv}. 

\subsection{Asymptotic freedom}
In asymptotically free theories 
with a trivial UV fixed point such as QCD, the canonical mass dimension of invariants in the fundamental action dictates whether the corresponding couplings are relevant, marginal, or irrelevant at highest energies. Then, standard dimensional analysis can be applied to conclude that operators with increasing canonical mass dimension will become increasingly irrelevant in the UV. Stated differently, for asymptotically free theories the set of universal eigenvalues
\begin{equation}
\label{thetaGauss}
\{\vartheta_{{\rm G},n}\}
\end{equation}
is known a priori, and given by the Gaussian values. The before-hand knowledge of the set  \eq{thetaGauss}, and, therefore, the fundamental action and its relevant or marginal free parameters,  is at the root for reliable approximation schemes for asymptotically free theories, eg.~those used in perturbative or lattice QCD.  If quantum Einstein gravity were asymptotically free, its Gaussian values would simply be given by \eq{classicalscaling}, modulo mulitplicities.

\subsection{Asymptotic safety}
In the absence of asymptotic freedom, residual interactions at highest energies become important. Quantum scale invariance can be achieved provided the theory develops
a non-trivial UV fixed point.
However, a perturbative operator ordering according to canonical mass dimension can no longer be taken for granted and  the set of relevant, marginal, and irrelevant operators will be modified.  Unlike for asymptotic free theories, and in the absence of further information about the nature and structure of these interactions, the set of universal eigenvalues at an asymptotically safe UV fixed point 
\begin{equation}\label{thetaUV}
\{\vartheta_n\} 
\end{equation}
is not known a priori. 
Any set of eigenvalues \eq{thetaUV}  whose  subset of negative eigenvalues remains finite  would be in accord with the  principles of the asymptotic safety conjecture. In turn, 
the fixed point theory could lose its predictive power if infinitely many eigenvalues changed their sign in the step from \eq{thetaGauss} to \eq{thetaUV} due to quantum corrections. 
We conclude that the feasibility of an asymptotic safety scenario necessitates that invariants
with a sufficiently large canonical mass dimension remain irrelevant even at an asymptotically safe UV fixed point  \cite{Weinberg:1980gg}.

\subsection{Bootstrap hypothesis}

The observation that an interacting quantum theory may, 
potentially, develop many ways to become asymptotically safe leads to a lack of a priori information about the relevancy or irrelevancy of operators and their eigenvalues \eq{thetaUV}. In practice, tests for asymptotic safety with lattice or continuum methods are often bound to  a finite set of invariants $\{{\cal O}_i\}$ retained in the fundamental action. If the theory displays RG fixed points, these necessarily will have finitely many relevant  eigendirections \eq{thetaUV}. How can we then be certain that this approximate study provides us with a reliable snapshot of the physical theory? We would need to know whether further invariants, eg.~some of those not retained in the study, will not lead to new relevant directions. This dilemma is by no means generic to asymptotic safety of gravity. This conceptual challenge arises whenever perturbatively non-renormalisable theories are tested for their non-perturbative renormalisability, including non-gravitational ones, eg.~non-linear $\sigma$-models and Gross-Neveu models in more than two space-time dimensions, and QCD in more than four space-time dimensions.

In \cite{Falls:2013bv}, we have proposed to circumnavigate this dilemma with the help of a bootstrap. The idea is to compensate, at least partly, the lack of a priori information for \eq{thetaUV}  by a working hypothesis for the operator ordering at an interacting fixed point. We will assume that
\begin{equation}\label{H1}
\bullet 
\begin{array}{rl}
&{\rm 
the\ relevancy\ of\ invariants\ at\ an\ interacting\ fixed\ point\ continues}\\ &{\rm to\ be\ governed\ by\ the\ invariant's\ canonical\ mass\ dimension}\,.
\end{array}
\end{equation}
The hypothesis  trivially holds true for any  non-interacting theory, and in particular for asymptotically free (UV) fixed points. It also holds true for theories with a weakly-coupled (UV) fixed point where anomalous dimensions of invariants are perturbatively small, see \cite{Litim:2014uca} and references therein. By continuity in the coupling strength, we expect that this persists even in the interacting theory, at least for invariants with a sufficiently large canonical mass dimension.  
This point of view relates with an observation made earlier in \cite{Weinberg:1980gg}: there, it has been argued to be unlikely that invariants with a large canonical mass dimension will become relevant at an asymptotically safe fixed point, because quantum corrections would have to be strong enough to revert the sign of increasingly large canonical mass dimensions. On the other hand, it is expected that low order eigenvalues become strongly modified, including changes of signs, as a consequence of interactions. 

The main benefit of a physically motivated working hypothesis such as \eq{H1} is that it can be put to the test by using the canonical mass dimension of invariants as the ordering principle \cite{Falls:2013bv}. If the hypothesis is confirmed from order to order in an expansion in the canonical mass dimension of invariants, this would strengthen the view that the fixed point is a stable property of the theory, even beyond those orders studied explicitly.

\subsection{Testing asymptotic safety}
\begin{figure}[t]
\centering
\begin{center}
\includegraphics[width=.7\hsize]{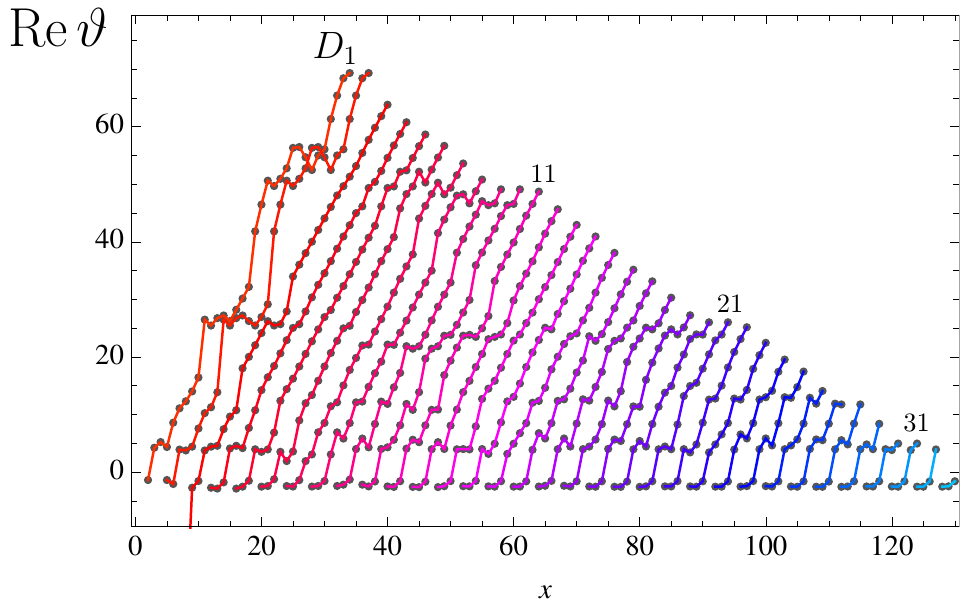}
\caption{\label{pBootstrap} The bootstrap test for asymptotic safety. From left to right, each line shows the entries of the $i^{\rm th}$ diagonal $D_i$  \eq{Di} of the eigenvalue matrix \eq{TnN}, with $i=1,\cdots, 33$.
The left-most line $D_1$ thus connects the largest eigenvalue at approximation order $N_{\rm max}$ with the largest at order $N_{\rm max}-1$, and so forth. 
The positive slope of all curves $D_i$ indicates that the working hypothesis is satisfied on average, although not for each and every order. 
  (see main text).}
\end{center}
\end{figure}

In Fig.~\ref{pBootstrap} we summarize the evidence in support of the working hypothesis \eq{H1}. We display the order-by-order variation of eigenvalues in the following manner. Each line $D_i$ in Fig.~\ref{pBootstrap} for $i=1$ to $33$, shows the eigenvalue set \eq{Di} -- the diagonals of the eigenvalue matrix $T$ introduced in \eq{TnN} -- thus showing the $i^{\rm th}$ largest eigenvalue from all approximation orders $N$ which have at least $N_{\rm max}+1-i$ eigenvalues. For example, the left-most line $D_1$ connects, from top right to bottom left, the largest eigenvalue at approximation order $N_{\rm max}$ with the largest at order $N_{\rm max}-1$, and so on, decreasing in steps of $\Delta N=\Delta x=1$.
The base points for the sets $D_i$ are located at $x(N_{\rm max})=31+3i$ for better display.
The working hypothesis states that the addition of an invariant with a new largest canonical mass dimension should result in the appearance of a new largest eigenvalue, larger than those encountered at lower orders in the approximation. If realised in the data, this pattern requires that all curves in   Fig.~\ref{pBootstrap}, on average, should rise  from order to order (with increasing $x$).  This is confirmed from the data: the positive slope of all curves $D_i$ indicates that the working hypothesis is satisfied. In particular for all curves from $D_3$ onwards this pattern is very stable, except for a few sideward variations, which occur precisely when a complex eigenvalue settles in the spectrum. Then, as discussed in Sect.~\ref{deg}, their real parts become degenerate. The stronger variation in the largest and second largest eigenvalue sets $D_1$ and $D_2$ can also be understood. These are related to the fact that the largest eigenvalues, more often than not, come out as  a complex conjugate pair. When this happens, as detailed in Sect.~\ref{CP}, these eigenvalues are often not reliable quantitatively, and the presence of more couplings is required before these start converging towards their asymptotic values. From the data, this already happens visibly from the set of third largest eigenvalues $D_3$ onwards. We conclude that the fixed point is self-consistent in the sense coined above.

\begin{figure}[t]
\begin{center}
\includegraphics[width=.7\hsize]{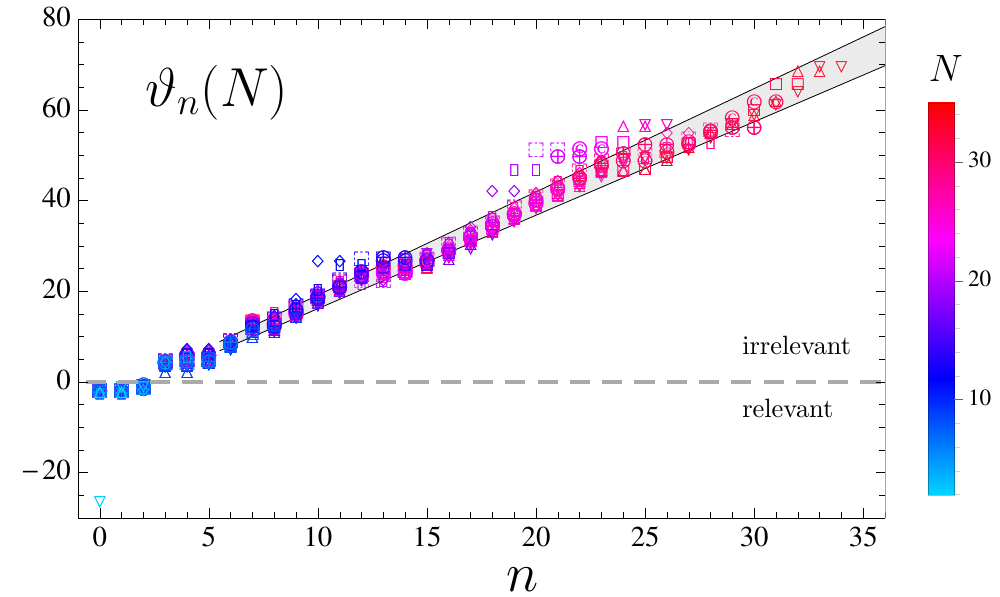}
\caption{\label{pAll}The overlay of all data sets for the universal scaling exponents $\vartheta_n(N)$ for $2\le N\le 35$ \cite{Falls:2013bv}, amended by the fit for the large-order behaviour \eq{fit}, \eq{abUV} including estimated errors (shaded area).}
\end{center}
\end{figure}

\section{Near-Gaussianity}\label{NG}
In this section, we discuss the large-order behaviour of universal eigenvalues.

\subsection{Large-order behaviour}
Expanding the analysis given in \cite{Falls:2013bv}, we show in Fig.~\ref{pAll} the sets of all eigenvalues from all approximation orders \eq{TnN} on top of each other. We find that the eigenvalues $\vartheta_n$ vary by about 20\% due to the inclusion of higher order invariants with $N>n+1$. As already noted earlier, the largest deviations from the best estimate $(N=35)$ arise from those lower-order approximations for which the largest eigenvalues are a complex conjugate pair. These, however, then stabilise rapidly with increasing approximation order.  Fig.~\ref{pAll} also confirms the good numerical convergence of exponents for all $n$. Most interestingly, we also observe that the real part of the asymptotically safe exponents become near-Gaussian  \cite{Falls:2013bv}.  To see this more quantitatively, we have performed in \cite{Falls:2013bv} a least-square linear fit of the real parts of the eigenvalues per approximation order in the form 
\begin{equation}\label{fit}
\vartheta_n= a\cdot n -b\,.
\end{equation}
for 24 data sets with $11\le N\le 34$. For each of these fits, we omit the two largest values for the reasons detailed earlier. We also omit the first few lowest exponents, as these may not yet display the large-$n$ asymptotics. We find that the correlation coefficients are very close to one for the fits of all data sets, supporting the applicability of the parametrisation \eq{fit}. We have also tested fits to higher polynomials in $n$, finding that the coefficients for the non-linear terms are negligible. The non-perturbative coefficients in \eq{fit} at the ultraviolet fixed point come out as  \cite{Falls:2013bv}
\begin{equation}
\label{abUV}
\begin{array}{rcl}
a_{\rm UV}&=&2.17\pm 5\% \\[.5ex]
b_{\rm UV}&=&4.06\pm10\%\,,
\end{array}
\end{equation}
where the error estimate, roughly a standard deviation, arises from the average over data sets  \cite{Litim:2010tt}.  Fig.~\ref{pAll} shows all data sets including the fit \eq{fit}, \eq{abUV} within its estimated errors, indicated by the shaded area.  
Classically, the universal eigenvalues would take Gaussian coefficients
\begin{equation}\label{GaussFit}
\begin{array}{rcl}
a_G&=&2\\[.5ex]
b_G&=&4\,.
\end{array}
\end{equation}
The differences between \eq{abUV} and the Gaussian coefficients \eq{GaussFit} serve as an indicator for the non-perturbative corrections due to asymptotically safe interactions. Our results 
establish that the UV scaling exponents remain near-Gaussian at high orders. 
The off-set $b_{\rm UV}$ is compatible with the classical value, though with a slight bias towards larger values, whereas the slope $a_{\rm UV}$ comes out larger than the Gaussian slope. 
It is tempting to speculate that this may be a consequence of the smallness of Newton's coupling at an ultraviolet fixed point. 

\begin{figure}[t]
\begin{center}
\includegraphics[width=.7\hsize]{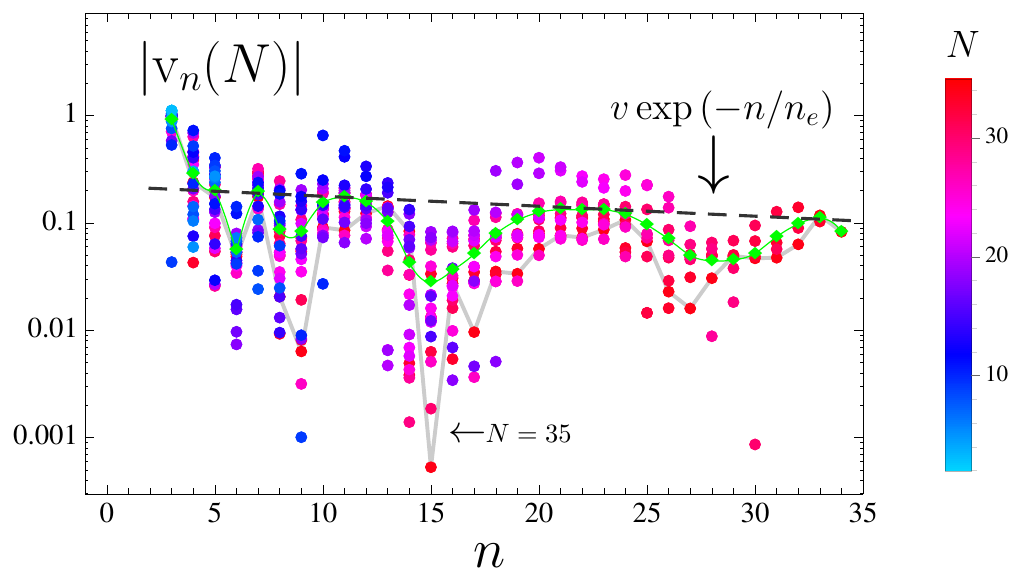}
\caption{\label{Relative}The relative variation \eq{Delta} of the non-perturbative eigenvalues $\vartheta_n(N)$ with respect to their Gaussian counterparts $\vartheta_{G,n}$, including data from all approximation orders $4\le N\le 35$. Mean values for each $n$ (green dots) are connected by a wavy full green line. A thin gray line connects the data at order $N=35$. With increasing $n$,  the envelope \eq{asym3} provides a good estimate for the upper bound (dashed line).}
\end{center}
\end{figure}

The smallness of the estimated error in the coefficients \eq{abUV}  has the additional benefit that it permits an extrapolation of the result \eq{fit} towards higher $n$. In particular, our results  indicate that even higher order invariants of the form $\int\sqrt g  R^{M-1}$ with $M>N_{\rm max}$ will only add increasingly irrelevant eigendirections at the UV fixed point. These observations also show that the search for asymptotically safe fixed points can reliably be limited to a finite polynomial basis of curvature invariants.

\subsection{Eigenvalue shifts}\label{ES}
The near-Gaussianity of large-order eigenvalues can be made more precise. In Fig.~\ref{Relative}, we show a semi-logarithmic plot for  the relative shift of the eigenvalues away from Gaussian values, introducing
\begin{equation}\label{Delta}
{\rm v}_n(N)=1-\frac{{\rm Re}\,\vartheta_n(N)}{\vartheta_{{\rm G},n}}\,.
\end{equation}
The colour-coding of the data shows the trend that $|{\rm v}_n(N)|$ decreases with increasing $N$. 
Based on the data up to $1/N_{\rm max}\approx 0.03$,
we conclude that \eq{Delta} resides in the  10--20\% range, 
\beq\label{asym2}
|{\rm v}_n(N)|<0.1-0.2\,,
\eeq 
decreasing with increasing $n$. In addition, we estimate the asymptotic behaviour of \eq{Delta} by taking the average values  for each $n$ over all approximation orders $N$. These are indicated in Fig.~\ref{Relative} by green dots and connected with a green line to guide the eye. The mean values  show a much smoother dependence on $n$, slowly decaying with increasing $n$. Their envelope is characterised by four maxima which are fitted very well by a simple exponential, 
\beq\label{asym3}
{\bar {\rm v}_n}\approx v\cdot \exp\left(-\frac{n}{n_e}\right)\,.
\eeq 
In Fig.~\ref{Relative}, the envelope of mean values \eq{asym3} is shown by a black dashed line. All mean values from $n>5$ onwards, and most entries from the high-order data sets, are below the envelope. Quantitatively, we have
\beq\label{asym4}
\begin{array}{rcl}
v&=&0.220\pm0.003\\[.5ex]
n_e&=&46.68\pm0.92\,.
\end{array}
\eeq 
The significance of  \eq{asym3} with \eq{asym4} is as follows. The parameter $v$ is a measure for the mean relative variation in \eq{Delta} at low $n$, and the parameter $n_e$ states at which order the relative variation becomes reduced by a factor of $e$. With $N_{\rm max}/n_e\approx \s034$, the reduction at $N_{\rm max}$ is by a factor of $\s012$, consistent with \eq{asym2}.
The new piece of information here is that the data shows a consistent, albeit slow, asymptotic decay towards near-Gaussian values. If this pattern persist to higher orders, extrapolation of \eq{asym3}, \eq{asym4}  predicts that 
\begin{equation}\label{vto0}
{\rm v}_n(N)\to 0
\end{equation}
for sufficiently large $n$, and
$1/N\to 0$.
This is interesting inasmuch as near-Gaussian eigenvalues are not mandatory for the asymptotic safety conjecture to apply.
For example, deviations such as \eq{asym2}, or even more substantial modifications of eigenvalues up to
\begin{equation}
{\rm v}_n(N)< 1
\end{equation}
at large orders would still be compatible with asymptotic safety. In this sense,  in our gravity model the quantum modifications of the high-order eigenvalues at the fixed point are moderate.  It is then conceivable that asymptotic safety persists under the inclusion of  further curvature invariants beyond those studied here. 

\begin{figure}[t]
\begin{center}
\includegraphics[width=.7\hsize]{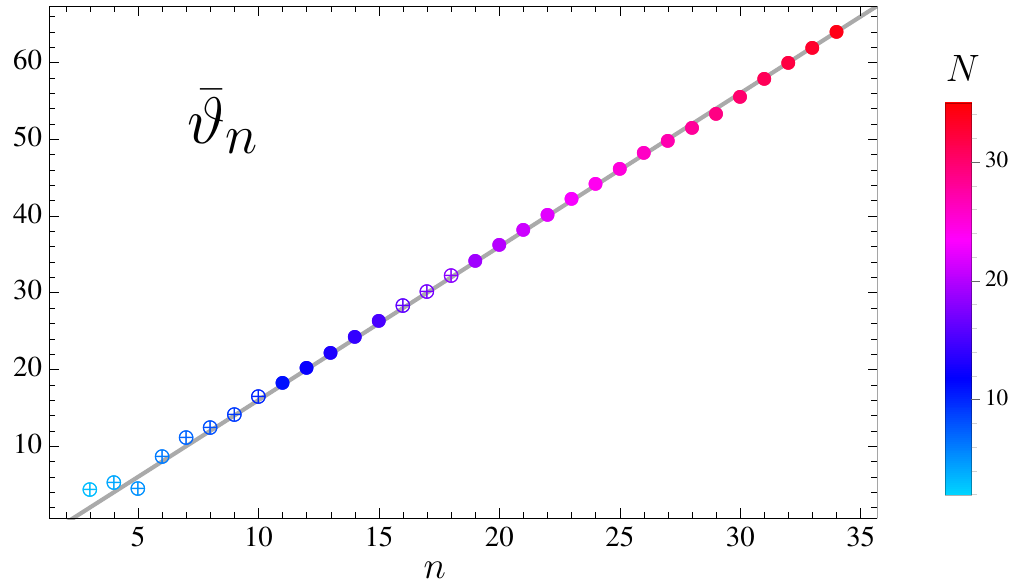}
\caption{\label{pMax}Shown is $\bar\vartheta_n$, the largest real eigenvalue at approximation order $N=n+1$, as a function of $n$ and
in comparison with the  Gaussian eigenvalues  $\vartheta_{G,n}$ (full line).  Crossed circles (full dots) indicate that $\bar\vartheta_{n}$ arises as the (third) largest real eigenvalue at polynomial approximation order $N=n+1$, see \eq{max}.
}
\end{center}
\end{figure}

\subsection{Origin of near-Gaussianity}
The appearance of near-Gaussian eigenvalues at large orders despite of a non-trivial, interacting, fixed point is quite intriguing. Here, we want to shed some light into its origin. 
In Figs.~\ref{AllEigenvalues} and \ref{Convergence} we already observed that, at fixed approximation order $N$, 
at least one of the three 
eigenvalues 
$\vartheta_{N-1}$, $\vartheta_{N-2}$, and $\vartheta_{N-3}$
 is real. The largest real eigenvalue is then either $\vartheta_{N-1}$ or $\vartheta_{N-3}$. 
If the eigenvalues with the largest real part are a complex conjugate pair $\vartheta_{N-1}=\vartheta^*_{N-2}$, their values
are numerically less reliable as these change visibly for approximation orders $>N$. 
On the other hand, if $\vartheta_{N-1}$ is real, it appears to only change mildly compared to approximation orders $>N$. 
Therefore one may suspect that the largest real eigenvalue within each set of eigenvalues \eq{ThetaN}  
is already a good estimate for the physical eigenvalue.  

Specifically, we wish to check whether the physical eigenvalue $\vartheta_{n}$ for large $n$ is already well-approximated 
by the largest real eigenvalue  at approximation order $N=n+1$, which is the lowest approximation order at which a non-zero 
eigenvalue $\vartheta_n(N)=\vartheta_{N-1}(N)$ arises in the spectrum. We denote this eigenvalue as
\beq\label{max}
\bar\vartheta_n=\max_m  \, \vartheta_m(N=n+1)\,\Big|_{{\rm Im}\,\vartheta_m=0}\,.
\eeq
Empirically, as already mentioned, $\bar\vartheta_n$  is then either $\vartheta_{N-1}$ or $\vartheta_{N-3}$
in the set of eigenvalues $T_N$ $(N=n+1)$; see \eq{ThetaN}. 
In Fig.~\ref{pMax} we display \eq{max} as a function of fixed $n=N-1$
from which it had been taken. Crossed circles indicate that the eigenvalue defined in \eq{max} is $\bar\vartheta_n=\vartheta_{N-1}$
of the  set $T_N$,
whereas full dots indicate $\bar\vartheta_n=\vartheta_{N-3}$
and hence the existence of a complex conjugate pair of eigenvalues with a larger real part. 

We first compare   $\bar\vartheta_n$  for different approximation orders $n=N-1$ with the Gaussian eigenvalues  $\vartheta_{G,n}$  \eq{classicalscaling}, shown
by the full line  in Fig.~\ref{pMax}. 
For low values of $n$, the largest real eigenvalue $\bar\vartheta_{n}$  differs slightly from $\vartheta_{G, n}$. For larger $n$, both lines are  on top of each other at the percent level and below, showing that
\beq\label{asym1}
\frac{\bar\vartheta_{n}}{\vartheta_{G, n}}\to 1
\eeq 
for $1/n\to 0$. Hence, all eigenvalues \eq{max} are near-Gaussian.  Next we keep $n$ fixed but increase the approximation order to $N_1>N=n+1$. 
We recall from the previous subsections that the results for $\vartheta_n(N_1)$ 
from high enough  approximation orders $N_1$ 
are 
also approaching near-Gaussian values, eg.~\eq{vto0}. Numerically, the inclusion of further operators results in a 10--20\% 
shift once the underlying higher-order couplings have settled. The extrapolations \eq{vto0} and \eq{asym1} of the full data 
shows that these deviations decrease even further,
\begin{equation}\label{limit}
\vartheta_n\approx \bar\vartheta_n
\end{equation}
for sufficiently large $n$, beyond $N_{\rm max}$ studied here. We conclude that $\bar\vartheta_n$ in \eq{max} is a good estimate for $\vartheta_n$, already
on the 10--20\% level for the approximations studied here, and, also in view of \eq{limit}, increasingly better with increasing $n$.

\subsection{``As Gaussian as it gets''}

We  close with a brief discussion of the main physics picture as it has emerged from our study.  We have analysed the effect of quantum fluctuations for a template version of 4D quantum gravity whose action is a high order polynomial in the scalar curvature. The primary effect of the quantum fluctuations of the metric field is to generate an interacting gravitational fixed point for all couplings. These effects have conveniently been parametrised in terms of a curvature-dependent function $f(R)$.\footnote{This is similar in spirit to studies of strongly-coupled QED${}_{d=4}$, where quantum effects have been parametrised in terms of a non-perturbative  anomalous dimension, e.g.~\cite{PhysRevD.42.3514,PhysRevD.45.4672}. The present model may equally be rewritten in terms of a curvature-dependent anomalous dimension for the graviton.} The gravitational fixed point comes out strongly coupled in the sense that the graviton anomalous dimension becomes large, of order unity. The fixed point is self-consistent in that it arises consistently, order by order in the polynomial approximation of the underlying action.   The fixed point is  physical in that gravity remains an attractive, albeit much weakened, force at highest energies.

The vacuum energy and Newton's constant remain relevant couplings in the UV even in the presence of quantum fluctuations, as one might have expected based on dimensional analysis. 
The classically marginal $R^2$ invariant becomes relevant quantum-mechanically.  
  Higher order interactions $R^n$ (with $n\ge 3$) all remain irrelevant in the UV, dynamically, despite of residual interactions.
The theory thus has  a three-dimensional UV critical surface. UV finite  trajectories emanating out of the fixed point are characterised by three  parameters, which must be viewed as free parameters of the fundamental theory. Ultimately, these are not fixed by the UV fixed point itself and can only be determined by experiment or observation.  

Quantitatively, on the level of the universal exponents, quantum effects induce a shift $\Delta \vartheta_n$ 
away from Gaussian values,
\beq\label{shiftexponents}
\vartheta_{G, n}\to\vartheta_{n}= \vartheta_{G, n}+\Delta \vartheta_n
\eeq
Most notably, with increasing canonical mass dimension of curvature invariants  we also observed that the universal scaling exponents \eq{shiftexponents} become ``nearly Gaussian'', as a consequence of 
\beq\label{smallness}
\Delta \vartheta_n/\vartheta_n\to 0\,,
\eeq 
with increasing $n$, see \eq{vto0}, \eq{asym1}.  The smallness of \eq{smallness} would seem to suggest that a small expansion parameter is hidden in the model. 
This result is  intriguing because the perturbative non-renormalisability of gravity disallows  an asymptotically free UV fixed point with exact Gaussian scaling. Instead, in the presence of residual UV quantum fluctuations,  the gravitational couplings must re-arrange themselves away from Gaussian values. Dynamically, they do this in such a manner that their universal scaling exponents remain nearly Gaussian.\footnote{Examples of non-gravitational quantum field theories with exact Gaussian scaling exponents at interacting fixed points  are known in lower dimensions, e.g.~$(\phi^2)^3_{d=3}$ at large-$N$.} From this point of view, the interacting theory has become ``as Gaussian as it gets''. The price to pay for the theory's perturbative non-renormalisability  is that its quantum theory  displays three relevant directions, rather than than  two relevant and a marginal one. No further relevant directions (and hence no new fundamentally free parameters) are induced by higher order curvature invariants $R^n$ once $n>2$. Still, the presence of higher order couplings is of importance on a quantitative level
inasmuch as they stabilise the fixed point for the lower order curvature invariants and the scaling exponents. This affects most notably the $R^2$ coupling which has a vanishing canonical mass dimension: here, the feedback from higher order interactions is crucial to stabilise the $R^2$ interaction.

\section{Conclusions}\label{C}

We have put forward a detailed systematic search for asymptotically safe fixed points in four-dimensional quantum gravity 
for actions which are high-order polynomials in the Ricci scalar
\cite{Falls:2013bv}.  
Evidence for asymptotic safety
 is found order by order in a polynomial expansion of the 
action up to including 34 powers in the Ricci scalar, corresponding to $N=35$ independent curvature invariants, 
thereby exceeding earlier investigations \cite{Codello:2007bd,Codello:2008vh,Bonanno:2010bt} by more than twenty powers in the curvature scalar. The $N\to\infty$ limit has also been performed for the first time.
Fixed points and scaling exponents are stable, and the results
 predict a three-dimensional critical surface of couplings with non-Gaussian exponents, and near-Gaussian scaling exponents related to  invariants with a large canonical mass dimension. 

Our  findings also show that quantum scale invariance of gravity in the UV can be tested self-consistently by means of a bootstrap \cite{Falls:2013bv}. Scaling exponents only deviate moderately from classical values, suggesting that a polynomial expansion is viable despite of the facts that neither an explicit  small expansion parameter has been identified, nor that the exact set of relevant couplings was known beforehand (Fig.~\ref{pBootstrap}). 
Also owing to the near-Gaussianity of results, it is  safe to assume that the canonical mass dimension of invariants controls the relevancy of operators  at an interacting fixed point. It will be interesting to  test this pattern for actions with more complicated curvature invariants such as Riemann and Ricci tensor invariants, which offer more sensitivity to the dynamics of the metric field \cite{inprep2016}.

We have also found  structural hints for the near-Gaussian behaviour of eigenvalues as shown in 
Fig.~\ref{pMax}. If this is a property of the full quantum theory, it may be feasible to identify a small parameter underneath the mechanism for asymptotic safety. This is left for future work. On the technical side, we have put forward powerful algebraic and numerical methods to find exact expressions for fixed point candidates. The technique is quite general, and can be exploited even beyond the models studied here.

Our work can be expanded in several directions. 
First and foremost, it is mandatory to study quantum gravity beyond the tensor and momentum structures retained here, possibly including  non-local invariants  \cite{Tomboulis:1997gg,Modesto:2014lga}.
It will also be important to study extensions of functional RG flows beyond the present levels of approximation.  Of particular interest is the disentanglement of background and fluctuation fields  \cite{Litim:1998nf}, as first quantified in \cite{Litim:2002ce,Litim:2002hj,Bridle:2013sra} for scalar and gauge theories. Some of this has recently  been implemented  in \cite{Folkerts:2011jz}, and for  Einstein-Hilbert gravity in  \cite{Donkin:2012ud,Christiansen:2012rx,Codello:2013fpa,
Dona:2013qba,Becker:2014qya}. More work is required to exploit this for the theories considered here. Equally interesting are recent ideas to exploit convexity properties of the gravitational action \cite{Falls:2014zba}, which may help to simplify the systematics.

\section*{Acknowledgements}
We thank Edouard Marchais for discussions. 
This work is supported by the Science Technology and Facilities
Council (STFC) under grant number ST/J000477/1 and ST/L000504/1. KN  acknowledges support by the A.S. Onassis Public Benefit Foundation grant F-ZG066/2010-2011.

\appendix

\section{Fluctuation-induced interactions}\label{AppA}
In this section, we provide the explicit RG equations adopted in this paper. 
We recall the dimensionless version of the RG flow \eq{df},
\beq\label{appeq1}
\partial_t f(\R) -2 \R f'(\R)+4 f(\R)=I[f](\R)\,.
\eeq
The RHS encodes the contributions from fluctuations and arises from the operator trace \eq{FRG} over all propagating  fields. It generically splits into several parts,
\beq\label{Iapp}
I[f](\R)=
I_0[f] (\R)+ \partial_t f'(\R)\cdot I_1[f] (\R)+ \partial_t f''(\R)\cdot I_2[f](\R)\,.
\eeq
The  additional flow terms proportional to $\partial_t f'(\R)$ and $\partial_t f''(\R)$ arise through the Wilsonian momentum cutoff $\partial_t R_k$, 
which we have chosen to depend on the background field. Furthermore, the terms $I_0[f](\R)$, $I_1[f](\R)$ and $I_2[f](\R)$ depend on $f(\R)$  
and its field derivatives $f'(\R)$,  $f''(\R)$ and  $f'''(\R)$. There are no flow terms $\partial_t f'''(\R)$ or higher because the momentum cutoff  
$R_k$ is proportional to the  second variation of the  action. A dependence on $f'''(\R)$  in $I_0[f]$ results completely from rewriting  $\partial_t F''(\bar\R)$ in dimensionless form. 
In the following expressions, we will suppress the argument $\R=\bar R/k^2$. 

All three terms $I_0[f]$, $I_1[f]$, $I_2[f]$  arise from tracing over the fluctuations of the metric field for which we have adopted a transverse traceless 
 decomposition. The term $I_0[f]$ also receives $f$-independent contributions from  the ghosts and from the Jacobians originating from the 
split of the metrical fluctuations into tensor, vector and scalar parts. To indicate the origin of the various contributions in the expressions below, 
we use  superscipts $T$, $V$, and $S$ to refer to the transverse traceless tensorial, vectorial, and scalar origin. The specific form of $I_0[f]$, $I_1[f]$,
 $I_2[f]$ depends on the gauge choice as in Sec.~7 of \cite{Codello:2008vh}) and on the regulator choice (with the optimized cutoff \cite{Litim:2000ci,Litim:2001up}).
With these considerations in mind, we write the various ingredients in \eq{appeq1} as
\bea
\label{I0}
I_0[f] &=& 
c\,\left[
\frac{P_c^V}{D_c^V}+\frac{P_c^S}{D_c^S} 
+ \frac{P_0^{T1}\cdot f' +  P_0^{T2}\cdot\R \cdot f''}{D^T}
+ \frac{P_0^{S1}\cdot f' + P_0^{S2}\cdot f'' +  P_0^{S3}\cdot \R \cdot f''' }{D^S}
\right]
\\
\label{I1}
I_1[f] &=& 
c\,\left[
\frac{P_1^T}{D^T}+ \frac{P_1^S}{D^S}
\right]
\\
\label{I2}
I_2[f]&=&c\, \frac{P_2^S}{D^S}\ .
\eea
In our conventions, the numerical prefactor reads $c=1/(24\pi)$. It arises from our normalisation factor $16\pi$ introduced in \eq{f}, divided by the volume of the unit $4$-sphere, $384\pi^2$. Note that the factor is irrelevant for the universal exponents at the fixed point. 
The first two terms in \eq{I0} arise from  the vector (V) and scalar (S) parts of the ghosts and Jacobians, while the third and fourth arise from the tensorial $(T)$ and scalar $(S)$ metric fluctuations, respectively. Both \eq{I1} and \eq{I2} only have contributions from the tensorial and scalar metric fluctuations. The various denominators appearing in \eq{I0}, \eq{I1} and \eq{I2} are given by the $f$-dependent terms
\bea
D^T[f]&=&3f-(\R-3)f'\\
D^S[f]&=&2 f + (3-2\R) f' + (3-\R)^2 f''
\,.\eea
and the $f$-independent terms
\bea
\label{DVc}
D_c^V&=& (4-\R )\\
\label{DSc}
D_c^S&=& (3-\R)\ . 
\eea
The various terms $P$ in the numerators  of \eq{I0}, \eq{I1} and \eq{I2} are polynomials in $\R$. They arise through the heat kernel expansion of the traces, and are given by
\begin{eqnarray}
P_c^V&=&\ \ \, \frac{607}{15}\R ^2  -24 \R  -144\\
P_c^S&=&\ \ \,  \frac{511}{30}\R ^2-12 \R -36\\
P_0^{T1} &=& \ \ \,  \frac{311}{756} \R ^3-\frac{1}{3}\R ^2-90 \R +240\\
P_0^{T2}& =& -\frac{311}{756} \R ^3+\frac{1}{6}\R ^2+30 \R -60\\
P_0^{S1}&=&\ \ \, \frac{37}{756} \R ^3+\frac{29}{15}\R ^2+18 \R +48\\
P_0^{S2} &=&-\frac{37}{756} \R ^4-\frac{29}{10}\R ^3-\frac{121}{5}\R ^2-12 \R +216\\
\label{PS30}
P_0^{S3} &=& \ \ \, \frac{181}{1680}\R ^4+\frac{29}{15}\R ^3+\frac{91}{10}\R ^2-54\\
P_1^T &=&\ \ \,  \frac{311}{1512}\R ^3-\frac{1}{12}\R ^2-15 \R +30\\
P_1^S &=&\ \ \, \frac{37}{1512}\R ^3+\frac{29}{60}\R ^2+3 \R +6\\
P_2^S&=&-\frac{181}{3360}\R ^4-\frac{29}{30}\R ^3-\frac{91}{20}\R ^2+27\ .
\end{eqnarray}
From the explicit expressions it is straightforward to confirm that $I_0[f]$ has homogeneity degree zero in $f$,
\beq
I_0[a\cdot\,f]=\,I_0[f]
\eeq 
for any factor $a\neq 0$, whereas $I_1[f]$ and $I_2[f]$ have homogeneity degree $-1$, $I_i[a\cdot f]=a^{-1}\,I_i[f]$ $(i=1,2)$. This establishes that the entire fluctuation-induced contribution $I[f]$ on the RHS of the flow equation  \eq{Iapp} has homogeneity degree zero.

At a fixed point, the flow equation becomes a third order differential equation for $f(\R)$. When resolved for $f'''(\R)$, the RHS contains algebraic denominators which vanish for specific $\R$. These  points are
\beq
\begin{array}{rcl}
\R&=&3\\[.5ex]
\R&=&4
\end{array}
\eeq
 due to the $f$-independent terms \eq{DVc} and \eq{DSc}. Furthermore, the prefactor $\R\cdot P^{S3}_0$ of $f'''$ in \eq{I0} given in \eq{PS30} vanishes for real $\R$ at 
\beq
\begin{array}{rcl}
\R&=&-9.99\,855\cdots\,,\\[.5ex]
\R&=&\ \ \,0\,,\\[.5ex]
\R&=&\ \ \,2.00\,648\cdots\,.
\end{array}
\eeq
The point $\R=0$ is uncritical for our purposes. The other points will require some fine-tuning to extend a well-defined fixed point solution from small fields to arbitrary large fields. Note that the existence of these requirements also relates to technical choices of our approximation. 

Finally, we also provide the defining equations for eigenperturbations at a non-trivial fixed point, as required for the study in Sec.~\ref{eigenperturbations}. We consider small perturbations $\delta f$ about  the fixed point solution $f=f_*$ with $\partial_t f=0$ to find the differential equation
\beq\label{deltadelta}
\Big (1-E_2[f]\Big)\,\partial_t\,\delta f
=\Big(-4+2R\,\partial_R+E_3[f]\Big) \,\delta f
\eeq
for the eigenperturbations $\delta f$  to linearised order.
Here, the $n$th order differential operators $E_n$ are given by
\bea\label{E2}
E_2&=&I_1[f]\cdot \partial_\R+\,I_2[f]\cdot \partial_\R^2\\[1ex]
\label{E3}
 E_3 &=& 
c\,\left[
\frac{P_0^{T1}\cdot \partial_\R  +  P_0^{T2} \cdot \R \cdot  \partial^2_\R}{D^T[f]}
+ \frac{P_0^{S1}\cdot  \partial_\R  + P_0^{S2}\cdot \partial^2_\R  +  P_0^{S3}\cdot  \R \cdot \partial^3_\R  }{D^S[f]}
\right.
\nonumber\\
&&
\quad -\frac{P_0^{T1}\cdot   f' +  P_0^{T2}\cdot  \R \cdot  f''}{(D^T[f])^2}\Big(3-(\R-3)\partial_\R\Big)
\nonumber\\
&&
\left.
\quad
- \frac{P_0^{S1}\cdot   f' + P_0^{S2}\cdot f'' +  P_0^{S3}\cdot  \R \cdot  f''' }{(D^S[f])^2}\Big(2+(3-2\R)\partial_\R+(3-\R)^2\partial_\R^2\Big)
\right]
\eea

\bibliographystyle{apsrev4-1}
\bibliography{FoR_final}
\end{document}